\documentclass{iopart}
\usepackage{iopams} 
\usepackage{setstack}
\usepackage{geometry}
\usepackage{color}
\usepackage{amssymb}
\usepackage{mathrsfs}

\pdfoutput=1

\usepackage{graphicx,epsfig}
\usepackage{bm}

\usepackage{braket}

\newcommand{\be}{\begin{equation}}
\newcommand{\ee}{\end{equation}}
\newcommand{\bea}{\begin{eqnarray}}
\newcommand{\eea}{\end{eqnarray}}
\newcommand{\nn}{\nonumber\\}
\newcommand\eps{\epsilon}
\newcommand\veps{\varepsilon}
\def\Z{\rm Z_{GGE}}
\def\fr#1{(\ref{#1})}

\def\limprime{\sideset{}{'}\lim}

\renewcommand{\Eref}[1]{Eq.~(\ref{#1})}

\begin{document}
\title[Quantum Quench in the Transverse Field Ising chain II]{Quantum
Quench in the Transverse Field Ising Chain II: Stationary State Properties} 
\author{Pasquale Calabrese$^1$, Fabian H.L. Essler$^2$, and Maurizio Fagotti$^{2}$}
\address{$^1$ Dipartimento di Fisica dell'Universit\`a di Pisa and INFN - Pisa 56127, Italy}
\address{$^2$ The Rudolf Peierls Centre for Theoretical Physics, University of Oxford -  Oxford OX1 3NP, United Kingdom}
\begin{abstract}
We consider the stationary state properties of the reduced density
matrix as well as spin-spin correlation functions
after a sudden quantum quench of the magnetic field in the
transverse field Ising chain. We demonstrate that stationary state
properties are described by a generalized Gibbs ensemble.
We discuss the approach to the stationary state at late times.
\end{abstract}

\maketitle

\section{Introduction}%

This is the second of two papers on the quench dynamics of the transverse
field Ising chain. In the first part of our work \cite{us1}, which in
the following we will refer to as ``paper I'', we focussed on the time
dependence of the longitudinal spin correlations. The present
manuscript gives a detailed account of properties in the
\emph{stationary state} at infinite times after the quench.
An important motivation for studying problems of nonequilibrium time
evolution in isolated quantum systems is provided by recent
experiments on trapped ultra-cold atomic gases
\cite{uc,tc-07,tetal-11,cetal-12,getal-11,kww-06}. These experiments
suggest that observables such as multi-point correlation functions 
generically relax to time independent values. Such a behaviour at
first appears quite surprising, because unitary time evolution
maintains the system in a pure state at all times. The resolution of
this apparent paradox is that in the thermodynamic limit, (finite)
\emph{subsystems} can and do display correlations characteristic of a
mixed state, namely the one obtained by tracing out the degrees of
freedom outside the subsystem itself. In physical terms this means
that the system acts as its own bath. An important question is how to
characterize the reduced density matrix describing the stationary
behaviour. Intuitively one may expect that conservation laws will play
an important role, which in turn poses the question whether quantum
integrable systems exhibit qualitatively different stationary
behaviours when compared to generic, nonintegrable ones. These issues
have been addressed in a number of recent works
\cite{rev,gg,rdo-08,cc-06,cc-07,caz-06,mw-07,cdeo-08,bs-08,fcm-08,ke-08,scc-09,roux-09,sfm-11,fm-10,bdkm-11,kla-06,bkl-10,bhc-10,gce-10,INT,bpgda-10,meden,rf-11,ccrss-11,cic-11,rs-11,zch-11,spr-11,mc-12,gm-12,ck-12,mc-12b,ghl-12,ksci-12}. 
Most of these studies are compatible with the widely held belief (see
e.g. \cite{rev} for a comprehensive summary) that the reduced density
matrix of any finite subsystem (which determines correlation functions
of all local observables within the subsystem) of an infinite system
can be described in terms of either an effective thermal (Gibbs)
distribution or a so-called generalized Gibbs ensemble (GGE)
\cite{gg}. It has been conjectured that the latter arises for integrable
models, while the former is obtained for generic systems. Evidence
supporting this view has been obtained in a number of examples
\cite{gg,rdo-08,cc-07,caz-06,mw-07,cdeo-08,bs-08,scc-09,sfm-11,fm-10,bdkm-11}.
On the other hand, several numerical studies
\cite{kla-06,bhc-10,gce-10,mc-12,gm-12} suggest that the full picture
may well be more complex.
Moreover, open questions remain even with regard
to the very existence of stationary states. For example, the order
parameter of certain mean-field models have been shown to
display persistent oscillations \cite{MF1,MF2,MF3,MF4,MF5,MF6}.  
Non-decaying oscillations have also been observed numerically \cite{bhc-10}
in some non-integrable one-dimensional systems. This has given rise to
the concept of ``weak thermalization'', which refers to a situation where
only time-averaged quantities are thermal.

We will show in this manuscript that for a quench of the magnetic
field in the transverse field Ising chain, the reduced density matrix is
described by a GGE. This establishes that \emph{any local observable} is
described by the GGE. However we will also show that while some
two-point observables approach their asymptotic values relatively
quickly, others will do so only after times exponentially large in the
separation between the two points. In practice this precludes
experimental or numerical detection of stationary behaviour for these
observables. 

\subsection{Quench protocol and observables}
\label{s:quench}%

In the following we focus on a {\it global quantum quench} of the
magnetic field in the Ising Hamiltonian  
\be
H(h)=-J\sum_{j=1}^L\Bigl[\sigma_j^x\sigma_{j+1}^x+h\sigma_j^z\Bigr]\, ,
\label{Hamiltonian}
\ee
where $\sigma_j^\alpha$ are the Pauli matrices at site $j$, we assume $J>0$, $h>0$ and we
impose periodic boundary conditions 
$\sigma_{L+1}^\alpha=\sigma^\alpha_1$. 
We assume that the many-body system is prepared in the
ground state $|\Psi_0\rangle$ of Hamiltonian $H(h_0)$. At time $t=0$
the field $h_0$ is changed instantaneously to a different value $h$ and
one then considers the unitary time evolution of the system
characterized by the new Hamiltonian $H(h)$, i.e. the initial state
$|\Psi_0\rangle$ evolves as
\be
|\Psi_0(t)\rangle=e^{-itH(h)}|\Psi_0\rangle.
\ee
The above protocol corresponds to an experimental situation
\cite{uc,tetal-11,cetal-12}, in which a system parameter has been
changed on a time scale that is small compared to all characteristic
time scales present in the system. 

In equilibrium at zero temperature (and in the thermodynamic limit)
the Ising model \fr{Hamiltonian} exhibits ferromagnetic ($h < 1$) and
paramagnetic ($h > 1$) phases, separated by a quantum critical point
at $h_c=1$. The order parameter for the corresponding quantum phase
transition is the ground state expectation value $\braket{\sigma^x_j}$.
The Hamiltonian (\ref{Hamiltonian}) can be diagonalized by a
Jordan-Wigner transformation, which maps the model to spinless fermions
with local annihilation operators $c_j$, followed by a Fourier
transform and, finally, a Bogoliubov transformation. In terms of the
momentum space Bogoliubov fermions $\alpha_k$ the Hamiltonian 
is diagonal:
\be
H(h)=\sum_{k}
\veps_h(k) {\alpha}^\dagger_{k}{\alpha}_{k}\,, \qquad
\veps_h(k)=2J\sqrt{1+h^2-2h\cos(k)}.
\label{Hbog}
\ee
Details and precise definitions are given in Appendix A of paper I.

In the following we focus on the one and two-point functions of the
order parameter 
\bea
\rho^x(t)&=&\frac{\langle\Psi_0(t)|\sigma^x_\ell|\Psi_0(t)\rangle}
{\langle\Psi_0(t)|\Psi_0(t)\rangle} ,\\
\rho^{xx}(\ell,t)&=&\frac{\langle\Psi_0(t)|\sigma^x_{j+\ell}\sigma^x_j
|\Psi_0(t)\rangle}{\langle\Psi_0(t)|\Psi_0(t)\rangle}, \qquad  \rho^{xx}_c(\ell,t)=\rho^{xx}(\ell,t)-(\rho^x(t))^2,
\eea
the transverse spin correlators 
\bea
\rho^z(t)=\frac{\langle\Psi_0(t)|\sigma^z_\ell|\Psi_0(t)\rangle}
{\langle\Psi_0(t)|\Psi_0(t)\rangle},
\qquad  \rho^{zz}_c(\ell,t)=\frac{\langle\Psi_0(t)|\sigma^z_{j+\ell}\sigma^z_j
|\Psi_0(t)\rangle}{\langle\Psi_0(t)|\Psi_0(t)\rangle}
-(\rho^z(t))^2,
\eea
and the reduced density matrix of subsystem $A$
\be
\rho_{\rm A}(t)={\rm Tr}_{\rm \bar{A}}|\Psi_0(t)\rangle\langle\Psi_0(t)|.
\ee
Here $A\cup\bar{A}$ is the entire system and in practice we will take
$A$ to consist of $\ell$ consecutive sites along the chain.
${\rm Tr}_{\rm \bar{A}}$ denotes a trace in the space of states
describing the lattice sites in $\bar{A}$.

Following paper I, we divide the time evolution of a two-point
function for a fixed distance $\ell$ between the operator insertions
into three regimes, which are determined by the propagation velocity
$v(k)=\frac{d\veps_h(k)}{dk}$
of elementary excitations of the post-quench Hamiltonian. For a given
final magnetic field $h$, the maximal propagation velocity is 
\be
v_{\rm max}= \max_{k\in [-\pi,\pi]} |\veps'_h(k)|= 2J\min[h,1]\,.
\ee
The three different regimes are:
\begin{itemize}
\item Short-times $v_{\rm max} t\ll \ell$. 
\item Intermediate times $v_{\rm max} t\sim \ell$. This regime is of
particular importance for both experiments and numerical computations.
A convenient way of describing this regime is to consider evolution
along a particular ``ray'' $\kappa\ell=v_{\rm max}t$ in space-time,
see Fig.~\ref{fig:STSL}.
In order to obtain an accurate description of the dynamics at a
particular point along this ray, one may then construct an asymptotic
expansion in the single variable $\ell$ around the 
{\it space-time scaling limit} $v_{\rm max}t,\ell\to\infty$, $\kappa$
fixed.
\begin{figure}[t]
\begin{center}
\includegraphics[width=0.75\textwidth]{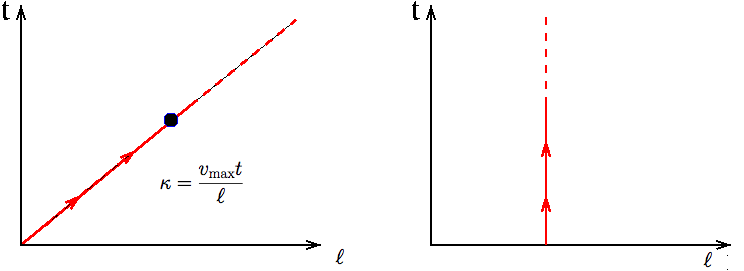}
\caption{Left panel: for intermediate times $v_{\rm max}t\sim\ell$ the
behaviour of $\rho^{\alpha\alpha}(\ell,t)$ is most conveniently
determined by considering its asymptotic expansion around infinity
(``space-time scaling limit'') along the ray $v_{\rm
  max}t=\kappa\ell$. This viewpoint is appropriate for any large,
finite $t$ and $\ell$. Right panel: the asymptotic late-time regime
is reached by considering time evolution at fixed $\ell$. To describe
this regime one should consider an asymptotic expansion of
$\rho^{\alpha\alpha}(\ell,t)$ around $t=\infty$ at fixed $\ell$.}
\label{fig:STSL}
\end{center}
 \end{figure}

\item Late times $v_{\rm max} t\gg \ell$. This includes the limit
  $t\to\infty$ at fixed but large $\ell$. In this regime it is no
longer convenient to consider evolution along a particular ray in
space-time. In order to obtain accurate results for the late time
dynamics, one should construct an asymptotic expansion in $t$
around infinity, see Fig.~\ref{fig:STSL}.
\end{itemize}
It is important to note that in general taking $\kappa\to\infty$ in
the space-time scaling limit does not necessarily reproduce the late
time behaviour at fixed, asymptotically large $\ell$. In other words,
in general we have 
\be
\limprime_{\kappa\to\infty}\
\limprime_{t,\ell\to\infty\atop \kappa\ {\rm
  fixed}}\ \rho^{\alpha\alpha}_c(\ell,t)\neq
\limprime_{\ell\to\infty}\
\limprime_{t\to\infty}\ \rho^{\alpha\alpha}_c(\ell,t),
\ee
where $\limprime$ denotes the leading term in an asymptotic expansion
around the limiting point.
We will show that the limits \emph{do commute} for $\rho^{xx}(\ell,t)$
for most but not all quenches, but they never commute for $\rho^{zz}_c(\ell,t)$.

\subsection{Longitudinal versus Transverse Correlators}

A global quantum quench of the transverse field in the Ising model is
an essentially ideal testing ground for many ideas related to
thermalization
\cite{rsms-08,can,CEF,mc,ir-00,ir-10,sps-04,fc-08,s-08,cz-10,fcg-11,ri-11,se12}. An example is the issue to what extent the late time
behaviour after a quench may depend on the observable under
consideration. In this regard, one may expect the \emph{locality} of
observables relative to the elementary excitations of the model to
play a role \cite{rsms-08,bhc-10,can}. In the Ising chain, the
transverse spin operator $\sigma^z_j$ is local with respect to the
fermionic degrees of freedom $c_j$, while the order parameter
$\sigma^x_j$ is in general non-local.
The non-locality makes the longitudinal correlators difficult to
analyze and general analytic results have been reported only recently
in our short communication \cite{CEF} (the stationary properties in
the special cases $h_0=0,\infty$ were obtained before in
Ref. \cite{sps-04}). 
The results reported in \cite{CEF} for the correlation length in the
stationary state were found to be described by an appropriately
defined GGE.

In contrast to order parameter correlators, the time evolution of one
and two point functions of the transverse spins $\sigma^z_m$ is
straightforward to analyze (as the latter are fermion bilinears) and
has been known since 1970 \cite{mc}. In spite of its simplicity, the
connected two-point function $\rho^{zz}_c(\ell,t)$ (corresponding to density
correlations in gases) exhibits an interesting and quite general
phenomenon, namely a cross-over in the relaxational behaviour at an
exponentially large time scale $ t_{\rm cross}\sim e^\ell$. As a consequence
the truly stationary regime of $\rho^{zz}_c(\ell,t)$ is observed
only at very late times $t>t_{\rm cross}$, which in general is a
serious limitation. 

\subsection{The quench variables}\label{s:The model}%

As shown in Appendix A of paper I, both the initial and final
Hamiltonians can be diagonalized by combined Jordan-Wigner
and Bogoliubov transformations with Bogoliubov angles
$\theta_k^0$ and $\theta_k$ respectively
\bea
e^{i\theta_k}=\frac{h-e^{ik}}{\sqrt{1+h^2-2h\cos k}}\ ,\qquad
e^{i\theta_k^0}=\frac{h_0-e^{ik}}{\sqrt{1+h_0^2-2h_0\cos k}}.
\label{Bangle}
\eea
The corresponding Bogoliubov fermions are related by a linear
transformation characterized by the difference
$\Delta_k=\theta_k-\theta^0_k$.  
In order to parametrize the quench it is useful to introduce
the quantity
\be
\cos \Delta_k=\frac{h h_0- (h+h_0) \cos k+1}{\sqrt{1+h^2-2h\cos(k)} \sqrt{1+h_0^2-2h_0\cos(k)}}.
\label{cosDeltak}
\ee
We note that $\cos \Delta_k$ is invariant under the two transformations
\bea
(h_0,h)&\rightarrow (h,h_0)\quad  {\rm and}\quad
(h_0,h)\rightarrow \Big(\frac{1}{h_0},\frac{1}{h}\Big)\ .
\label{maps}
\eea
However, we stress that the quantum quench itself is not invariant
under the maps \fr{maps}. 

\subsection{Generalized Gibbs ensemble}

The density matrix of a GGE can be cast in the form \cite{GGE}
\be
\rho_{\rm GGE}=\frac{1}{{\rm Z}_{\rm GGE}} e^{-\sum_m \lambda_m I_m}\,, 
\label{ggedef0}
\ee
where $I_m$ are some integrals of motion and $\Z$ ensures the
normalization condition $\Tr \rho_{\rm GGE}=1$. 
In Ref. \cite{gg}, Rigol, Dunjko, Yurovsky and Olshanii
proposed that an integrable system after a quantum quench in the
infinite time limit is in fact described by a GGE, where   
the $I_m$ represent a complete set of independent integrals of motion
and the Lagrange multipliers $\lambda_m$ are fully determined by the
initial state $|\Psi_0\rangle$ through the conditions 
\be
\Tr [\rho_{\rm GGE} I_m]=\langle \Psi_0|I_m | \Psi_0\rangle\,.
\label{gge1}
\ee
As we are particularly interested in the thermodynamic limit, we will
employ the following, somewhat narrower, definition of a GGE for the
description of stationary properties after quantum quenches.
\begin{itemize}
\item{}
First, as the entire system will be in a pure state at all times, it
cannot be described by a density matrix corresponding to a mixed state.
We therefore take $\rho_{\rm GGE}$ to be the \emph{reduced density matrix}
of an infinitely large subsystem $A$ in the thermodynamic limit. 
From this reduced density matrix all multipoint correlation
functions can be obtained. Moreover, one can extract the reduced
density matrix of any finite subsystem $A_1$ by \cite{cdeo-08,bs-08}
\be
\rho_{A_1}(t=\infty)=\Tr_{\bar A_1} \rho_{\rm GGE}\, .
\ee 
Considering an infinitely large subsystem $A$ has the advantage that
the density matrix defining the GGE can be expressed in a simple way 
even for more complicated, interacting integrable models, whereas the
corresponding formulation for a finite subsystem becomes difficult.
However, in cases where the subsystem $A_1$ consists of several
disjoint blocks, the reduced density matrix will not have a simple
form like \fr{eq:GGE} even in the absence of interactions \cite{fc-10}.

The precise sequence of limits we have in mind is the following. The
entire system is decomposed in a subsystem $A$ and its complement
$\bar{A}$. We then take the thermodynamic limit, keeping $A$
fixed. Finally we take the limit of $A$ becoming infinite itself.
The operator $\rho_{\rm GGE}$ is the reduced density matrix of
subsystem $A$ in this limit.

The importance of considering the reduced density matrix when applying
GGE ideas has been emphasized previously in Refs. \cite{cdeo-08,bs-08,scc-09,bhc-10,gce-10}. 

\item{}
Another crucial point, that perhaps has not yet received the attention
it deserves, concerns the issue of what integrals of motion $I_m$ ought to be
included in the definition (\ref{gge1}) of the GGE density matrix.
Indeed, any quantum system, integrable or not, has many integrals of 
motion. For example the projectors $I_n=|\psi_n\rangle\langle \psi_n|$ on
Hamiltonian eigenstates $H|\psi_n\rangle =E_n|\psi_n\rangle$ are
integrals of motion $[H,I_n]=[I_m,I_n]=0$. Clearly, such conservation
laws cannot play any role in determining the late time behaviour after
a quantum quench. Following \cite{EF12} we therefore propose to
include only \emph{local} integrals of motion in (\ref{gge1}).
These are characterized by arising from an integral
(or a sum in the case of lattice models) of a local current density
$J_n(x)$ as $I_n=\int dx J_n(x)$, in the same spirit of the well known 
Noether theorem in quantum field theory.
This locality requirement is similar in spirit to the additivity
requirement of Ref. \cite{rigol11}.

\end{itemize}

With these remarks in mind,
the generalized Gibbs ensemble for the Ising chain is then defined by the
density matrix \cite{gg}
\be\label{eq:GGE}
\rho_{\rm GGE}=\frac{1}{\rm Z_{GGE}}{\exp\left({-\sum_k \beta_k \veps_k
    \alpha_k^\dag\alpha_k}\right)}\, , 
\ee
and the Lagrange multiplier $\beta_k$ are fixed by the initial state 
$|\Psi_0\rangle$ through the equations
\be
\langle\Psi_0|\alpha_k^\dag\alpha_k |\Psi_0\rangle=
\Tr [\rho_{\rm GGE} \alpha_k^\dag\alpha_k] \,.
\label{lagr}
\ee
For a quench of the magnetic field in the Ising chain,
these are easily solved with the results
\be
\beta_k\veps_k=2\mathrm{arctanh}(\cos\Delta_k)\ .
\label{gge2}
\ee
However, the integrals of motion $n_k=\alpha_k^\dag\alpha_k$ used in
\fr{eq:GGE} are \emph{non-local} in space. A priori this is a serious
problem as the integrals of motion defining a GGE must be local, as we
pointed out above. However, as shown in Ref. \cite{EF12}, the
prescription \fr{eq:GGE} works, because in the particular case of the
Ising model the local integrals of motion  can be expressed as
complicated {\it   linear} combinations of
$n_k=\alpha_k^\dag\alpha_k$. Hence for an infinite subsystem
\fr{eq:GGE} is equivalent to a GGE defined using the local integrals
of motion. 

\subsection{Organization of the manuscript.}

The present manuscript is organized as follows. In section
\ref{secsum} we give a detailed summary of our main results. In
Sec. \ref{sec:GGE} we show that the reduced density matrix is
described by a GGE. In Sec. \ref{sec:long} we compute the stationary
values of the longitudinal two-point correlation function, while in
Sec. \ref{sec:tran} we study transverse correlations and their
approach to the GGE. In \ref{szapp} we review some mathematical
theorems we need to evaluate the asymptotic behaviour of longitudinal
correlation functions.   

\section{Summary of Results}
\label{secsum}

\subsection{Reduced Density Matrix in the Stationary State}
Given that we start the system out in a pure state $|\Psi_0\rangle$,
its full density matrix is the one-dimensional projector
\be
\rho(t)=|\Psi_0(t)\rangle\langle\Psi_0(t)|.
\ee
The reduced density matrix of a subsystem $A$ is then obtained as
\be
\rho_A(t)={\rm Tr}_{\bar{A}}\big(\rho(t)\big),
\ee
where $\bar{A}$ is the complement of $A$. We have obtained the
following result for $\rho_{A}(t=\infty)$: in the thermodynamic limit,
for the case where $A$ is itself infinite, the reduced density matrix
in the stationary state is equivalent to that of the generalized Gibbs
ensemble \fr{eq:GGE}
\be
\rho_{A}(t=\infty)=\rho_{\rm GGE}.
\ee
In particular, this implies that arbitrary \emph{local} multi-point
spin correlation functions within subsystem $A$ can be evaluated as
averages within the GGE. However, the calculation of such averages
represents itself a formidable problem. We have determined these 
averages for the most important cases of the two-point functions 
$\rho^{xx}(\ell,t=\infty)$ and $\rho^{zz}(\ell,t=\infty)$ and present
explicit expressions below.
We stress that although $\sigma^x_n$ is not local in terms of fermions,
correlation functions involving $\sigma^x_n$ can nevertheless be
calculated from $\rho_{\rm GGE}$. For correlation functions involving
an even number of $\sigma^x_n$'s this is because the Jordan-Wigner
strings cancel and one is left with an expression that involves only
fermions within subsystem $A$. Correlation functions involving an odd
number of $\sigma^x_n$'s vanish as a result of the $\mathbb{Z}_2$
symmetry, which is restored in the GGE.

\subsection{Longitudinal Correlators in the Stationary State}
We find that the infinite time limit of the two-point function
$\lim_{t\to\infty}\rho^{xx}(\ell,t)$ for fixed but asymptotically
large $\ell$ is given by
\be
\rho^{xx}(\ell\gg1,t=\infty)= C^x(\ell)
e^{-\ell/\xi}\left[1+o\big(\ell^0\big)\right]. 
\label{rhoexp}
\ee
Here $\xi$ and $C^x(\ell)$ are functions of the initial ($h_0$) and
final ($h$) magnetic fields. The inverse correlation length is
\begin{eqnarray}
\fl  \xi^{-1}=\theta_H(h-1)\theta_H(h_0-1)\ln\left[\min(h_0,h_1)\right]-
\ln\Bigl[x_++x_-+\theta_H((h-1)(h_0-1))\sqrt{4 x_+x_-}\Bigr]\, ,
\label{xiasy}
\end{eqnarray}
where $\theta_H(x)$ is the Heaviside step function and
\be\fl \qquad\label{eq:h1}
x_\pm=\frac{[{\rm min}(h,h^{-1})\pm 1][{\rm min}(h_0,h_0^{-1})\pm 1]}4
\,, \quad  
h_1=\frac{1+h h_0+\sqrt{(h^2-1)(h_0^2-1)}}{h+h_0}\, .
\ee
The function $C^x(\ell)$ describes the subleading large-$\ell$
asymptotics and takes the following form
\begin{enumerate}
\item{} Quench within the ferromagnetic phase ($h_0,h<1$).
\be
C^x(\ell)=
\frac{1-h h_0+\sqrt{(1-h^2)(1-h_0^2)}}{2\sqrt{1-h 
    h_0}\sqrt[4]{1-h_0^2}}\equiv {\cal C}^x_{\rm FF}.
\ee
\item{} Quench from the ferromagnetic to the paramagnetic phase
($h_0<1<h$). 
\be
C^x(\ell)=
\sqrt{\frac{h\sqrt{1-h_0^2}}{h+h_0}}\equiv {\cal C}^x_{\rm FP}.
\ee
\item{} Quench from the paramagnetic to the ferromagnetic phase
($h_0>1>h$). 
\be
C^x(\ell)=
 \sqrt{\frac{\ h_0-h}{\sqrt{h_0^2-1}}} \cos\bigl(\ell
 \arctan\frac{\sqrt{(1-h^2)(h_0^2-1)}}{1+h_0 h}\bigr)
\equiv{\cal C}^x_{\rm PF}(\ell).
\label{CPFcos}
\ee
\item{} Quench within the paramagnetic phase ($1<h_0,h$). 
\be
\fl\qquad
C^x(\ell)\equiv{\cal C}^x_{\rm PP}(\ell)=
\left\{
\begin{array}{ll}
-\frac{h_0\sqrt{h}\bigl(h h_0-1+\sqrt{(h^2-1)(h_0^2-1)}\bigr)^2}{4
  \sqrt{\pi } (h_0^2-1)^{3/4} (h_0 h-1)^{3/2}(h-h_0)}\ \ell^{-3/2}
&{\rm if}\  1<h_0<h\ ,\\
 \sqrt{\frac{h(h_0-h)\sqrt{h_0^2-1}}{(h+h_0)(h h_0-1)}}&{\rm if}\  1<h<h_0.
\end{array}
\right.
\ee

\end{enumerate}
Here $\arctan(|x|)\in[0,\pi/2)$. We see that in most cases $C_x(\ell)$
tends to a constant value at large $\ell$. The exceptions are quenches
from the paramagnetic to the ferromagnetic phase, where $C^x(\ell)$
tends to an oscillatory function with constant amplitude, and quenches
to a larger magnetic field within the paramagnetic phase, where
$C^x(\ell)$ decays like a power law with exponent $-3/2$.
For the special cases $h_0=0$ and $h_0=\infty$ the asymptotic
behaviour of $\rho^{xx}(\ell\gg 1,t=\infty)$ agrees with the results
of Ref. \cite{sps-04}, which were obtained by different methods. 

\subsection{Transverse Correlators in the Stationary State}
The stationary behaviour of the connected longitudinal two-point
function is
\be
\rho^{zz}_c(\ell\gg1,t=\infty) \simeq C^z\ell^{-\alpha^z} e^{-\ell/\xi_z}(1+O(\ell^{-1}))\,,
\label{rhozzasy}
\ee
where the transverse correlation length $\xi_z$, the exponent
$\alpha^z$ and the amplitude $C^z$ are given by
\bea
\fl
\xi_z^{-1}&=&|\ln h_0|+\min(|\ln h_0|,|\ln h|),\nn
\fl\alpha^z&=&\left\{
\begin{array}{lcl}
1&&{\rm if\ \ } |\ln h|>|\ln h_0|\ ,\\
0&& {\rm if\ \ } h_0=1/h\ ,\\
1/2&&{\rm if\ \ } |\ln h|<|\ln h_0|\, ,
\end{array}
\right. \\
\fl C^z&=&\left\{
\begin{array}{lcl}
\frac{|h_0-1/h_0|}{4\pi}\frac{h_0-h}{h h_0-1}&&
{\rm if\ \ } |\ln h|>|\ln h_0|\ ,\\
-\frac{(h-1/h)^2}{2\pi} &&{\rm if\ \ } h_0=1/h\ ,\\
\frac{(h-1/h)\sqrt{|h_0-1/h_0|}(h_0-h)}{8\sqrt{\pi}
  h}\sqrt{\frac{h_0-h}{h_0(h h_0-1)}}\frac{e^{\mathrm{sgn}(\ln h)|\ln
    h_0|/2}}{\sinh\frac{|\ln h|+|\ln h_0|}{2}}&&{\rm if\ \ } |\ln
h|<|\ln h_0|\, .  
\end{array}
\right. 
\eea
\subsection{How long one has to wait before one can ``see'' the GGE}
Given that the infinite time behaviour of the reduced density matrix is
described by a GGE, a natural question to ask is how long we need to
wait in order to be able to detect the convergence of a given observable
to its stationary value. Clearly, to answer this question we
require knowledge of the full time evolution of correlation functions
and not only their stationary values. This is very difficult in
general, but can be studied in detail for the simple yet
instructive case of the transverse spin-spin correlation function. 
\subsubsection{Transverse Correlations}
\begin{figure}[t]
\begin{center}
\includegraphics[width=0.7\textwidth]{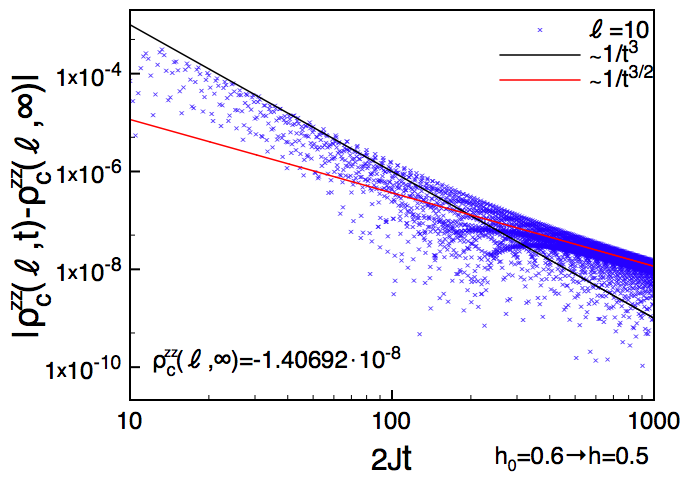}
\caption{Absolute value of the connected transverse correlator 
$\rho_c^{zz}(\ell,t)-\rho_c^{zz}(\ell,t=\infty)$ as
function of time at fixed $\ell=10$. Initial and final magnetic fields
$h_0=0.6\rightarrow h=0.5$ are chosen such that a crossover between a
$t^{-3}$ and the asymptotic $t^{-3/2}$ regime occurs at accessible
times.
}
\label{fig:transverse}
\end{center}
\end{figure}

The connected transverse spin-spin correlation function has the
following integral representation
\bea
\fl\quad
\label{eq:doubletrans_summary}
\rho_c^{zz}(\ell,t)
&=\int_{-\pi}^\pi\frac{\mathrm d k_1\mathrm d k_2}{(2\pi)^2}
e^{i \ell (k_1-k_2)}\Bigg\{\prod_{j=1}^2\sin\Delta_{k_j}
\sin\big(2\veps_h(k_j)t\big)\nn
\fl&\qquad\qquad\qquad\qquad-
\prod_{j=1}^2e^{i\theta_{k_j}}\Big[\cos\Delta_{k_j}-i\sin\Delta_{k_j}\cos\big(2\veps_h(k_j)t\big)\Big]\Bigg\}\, .
\eea
The integral representation can now be used to obtain asymptotic
expansions for large $\ell$ and $t$ in the two limits of interest:
\begin{enumerate}
\item{} In the space-time scaling limit $\ell,v_{\rm max}t\to\infty$
with $v_{\rm max}t/\ell$ fixed, the asymptotic behaviour can be
evaluated by means of a stationary phase approximation. This shows
that the leading behaviour is a $t^{-1}$ power-law decay, while
subleading corrections are power laws as well, i.e.
\be
\rho_c^{zz}\big(\ell= \frac{v_{\rm max}t}{\kappa},t\big)
\sim\frac{D^z(t)}{\kappa^2t}+o\big(t^{-1}\big)\ ,
\label{STSR}
\ee
where $D^z(t)$ is sum of a constant contribution and oscillatory terms
with constant amplitudes. 
\item{} In the late time regime at fixed, large $\ell$, we find that
$\rho_c^{zz}(\ell,t)$ decays as a power law in $t$ to a stationary
  value that is exponentially small in $\ell$
\be
\rho^{zz}_c(\ell,t)\sim 
\rho^{zz}_c(\ell,\infty)+\frac{E^z(t) \ell
  e^{-\ell/\tilde\xi_z}}{t^{3/2}}
+o\big(t^{-3/2}\big)\ ,
\label{t_to_infty}
\ee
where $E^z(t)=\sum_{q=0,\pi}A_q\cos\big(2t\veps_h(q)+\varphi_q\big)$
with constants $A_q$, $\varphi_q$, and the stationary value
$\rho^{zz}_c(\ell,\infty)$ is given above in \fr{rhozzasy}.
Crucially, $\rho^{zz}_c(\ell,\infty)\propto e^{-\ell/\xi_z}$
is exponentially small in $\ell$. The inverse correlation length
$\tilde \xi_z^{-1}$ is given by 
\be
\tilde \xi_z^{-1}=\min(|\log h_0|,|\log h|)<\xi_z^{-1}\, .
\ee
\end{enumerate}
We note that in the space-time scaling limit
exponentially small terms such as $\rho^{zz}_c(\ell,\infty)$ will
always be negligible compared to the dominant power law behaviour. On
the other hand, in the late time regime there exists a cross-over time
scale 
\be
Jt^*_2\sim e^{(2\ell/3)|\log h_0|},
\ee
after which the stationary behaviour becomes apparent. Importantly,
this time scale is \emph{exponentially large} in the separation
$\ell$. As we have alluded to before, the space-time scaling limit is
a convenient way of obtaining the behaviour of $\rho_c^{zz}(\ell,t)$
for general large $\ell$ and $t$. Contributions that are exponentially
small in $\ell$ are not included in the asymptotic expansion obtained
in the space-time scaling limit. This observation allows us to define
a cross-over time scale $t^*_1$ between the intermediate and late-time
regimes by the requirement that the leading asymptotics in the
space-time scaling limit ($\propto\ell^2/t^3$) becomes comparable to
the neglected exponentially small terms $O(e^{-\ell/\xi_z},e^{-\ell/\tilde\xi_z}/t^{3/2})$. This gives 
\be
Jt^*_1\sim 
e^{2\ell/3\tilde \xi_z}.
\ee
This estimate suggests that the result obtained in the space-time
scaling limit will give a good description of $\rho_c^{zz}(\ell,t)$
for \emph{fixed} separation $\ell$ up to times of order $t^*_1$.
When $|\log h_0|<|\log h|$, the two time scales are comparable
$t^*_1\sim t^*_2$. This means that in practice the space-time scaling
limit provides a good approximation for $\rho_c^{zz}(\ell,t)$ all the
way up to the stationary regime (see Fig. \ref{fig:transverse}). On the other hand, when
$|\log h_0|>|\log h|$ there is a large time window $t_2^*> t >
t_1^\ast$ in which a $t^{-3/2}$ power-law decay can be observed.

The implications of these considerations for experiments or numerical
calculations on finite-size systems are as follows:
\begin{itemize}
\item{} First and foremost, in order to observe the GGE either
numerically or experimentally, it is essential to focus on a suitable
observable. For the case of the TFIC this would be e.g. the one-point
function $\rho^z(t)$.
\item{} In a finite system of size $L$, boundary effects (such as
reflection from the boundaries or completing a full system traverse
in the periodic case) will become important at a timescale
\be
t_{\rm fs} \sim L/v_{\rm max}.
\ee
In order to observe stationary behaviour, i.e. the GGE, this timescale
should be larger than the cross-over scale $t_2^*$
\be
t_{\rm fs} > t^*_2.
\ee
In practice this requires extremely large system sizes. Just to quote
some numbers: if we could simulate chains of length $L=10000$ 
(which is at least one order of magnitude larger than what can be
achieved with currently available numerical algorithms), the GGE can
be ``seen'' at best for correlations evaluated at distances of about
$10$ lattice sites. Correlation functions evaluated at larger
distances may \emph{appear} stationary, but this behaviour will not be
related to the GGE.
\item{} The behaviour for large $v_{\rm max}t<v_{\rm max}t_{\rm fs}$
and distances $\ell$ (compared to the lattice spacing) for a finite
system will generically be described by the intermediate time regime
and in particular the result obtained in the space-time scaling limit.
Time scales necessary to observe stationary behaviour will generically
be inaccessible.
\end{itemize}

\subsubsection{Longitudinal Correlations}

As a result of their non-local nature with respect to the
Jordan-Wigner fermions these are much more difficult to determine than
the transverse correlations. Detailed results for the
time evolution of one and two-point functions of $\sigma^x_m$ are
presented in paper I. The implications of these results for the
present discussion are summarized as follows. 
\begin{itemize}
\item{} 
For quenches originating in the ferromagnetic phase, the stationary value of $\rho^{xx}(\ell,t)$ emerges after a time $t_3^*$ that scales 
as a power-law with the spatial separation $v_{\rm max}t_3^*\sim\ell^{4/3}$, {\it cf.} Eq. (28) of paper I. Hence it is straightforward to see the approach (as a function
of time) of the two-point function to its stationary values, which are
given by the GGE.
\item{} For quenches within the paramagnetic phase 
$\rho^{xx}_c(\ell,t)$ exhibits an oscillatory power-law decay in time
towards its stationary value, which is exponentially small in
$\ell$. Hence, in complete analogy to the case of the transverse
two-point function, the time scale $t^*_2$ after which the stationary
behaviour reveals itself, is exponentially large 
\be
t^*_2\propto e^{2\ell/3\xi},
\ee
and very difficult to observe in practice.
\end{itemize}

\subsection{Generalized Gibbs Ensemble and Conformal Field Theory.}

In Refs. \cite{cc-06,cc-07} it has been shown that for conformal
field theories the long time limit of correlation functions after a
quantum quench described by a conformally invariant boundary state
is described by a \emph{thermal} (Gibbs) ensemble.
Since CFTs are prototypical integrable field theories, this result
apparently contradicts the general expectation that integrable models
should be described by appropriately defined GGEs. 
We now show that for the case of the Ising field theory
there is in fact no contradiction and that the peculiarity of the CFT
result can be traced back to the particular (non-generic) initial
state considered in Refs. \cite{cc-06,cc-07}. 

As discussed in paper I, CFT describes the scaling limit of the Ising
model when the final gap is send to zero. The scaling limit of the
transverse field Ising chain is ($a_0$ is the lattice spacing) 
\be
J\to\infty\ ,\quad h\to 1\ ,\quad a_0\to 0,
\label{scalingI}
\ee
while keeping fixed both the gap $\Delta$ and the velocity $v$
\be
2J|1-h|=\Delta\ ,\quad 2Ja_0=v.
\label{scalingII}
\ee
In this limit the dispersion and Bogoliubov angle become
\bea
\veps(q)&=&\sqrt{\Delta^2+v^2q^2}\ ,\qquad
\theta_h(q)\rightarrow\mathrm{sgn}(h-1){\rm arctan}\Big(\frac{vq}{\Delta}\Big).
\eea
Here the physical momentum is defined as $q={k}/{a_0}$, with $ -\infty<q<\infty$.
In our quench problem both the initial and the final magnetic field
are scaled to the critical point, i.e. we need to take
\be
h_0\to 1\ ,\quad 2J|1-h_0|=\Delta_0=\ {\rm fixed}.
\ee
Thus in the scaling limit we have 
\be
\cos \Delta_k\to \frac{\Delta\Delta_0+(vq)^2}{\sqrt{(\Delta^2+(vq)^2)(\Delta_0^2+(vq)^2)}}\,.
\ee
For quenches to the quantum critical point, which is described by the
Ising conformal field theory with central charge $c=1/2$, we need to
furthermore take $\Delta\to 0$, which results in
\be
\cos \Delta_k\to \frac{v|q|}{\sqrt{\Delta_0^2+(vq)^2}}\,.
\ee
In this limit the expression (\ref{gge2}) for the Lagrange multipliers
$\beta_k$ becomes 
\be
\beta_q v |q|=2\ \mathrm{arctanh} \frac{v|q|}{\sqrt{\Delta_0^2+(vq)^2}}\, .
\label{TCFT}
\ee
This shows that one obtains mode-dependent temperatures and hence a
non-trivial GGE even in the case of a quench to the Ising CFT.

However, the particular boundary states considered as initial states in
Refs. \cite{cc-06,cc-07} correspond to the limit of infinitesimal
correlation length, i.e. a very large initial gap $\Delta_0\to\infty$.  
In the limit $v|q|/\Delta_0\to 0$, \fr{TCFT} becomes
\be
\beta_q=\frac{2}{\Delta_0}\,,
\ee
i.e. all mode dependent temperatures \emph{become equal}, resulting in
a thermal state. Thus the findings \cite{cc-06,cc-07} should not be
interpreted as showing that CFTs thermalize after quantum quenches.
We have been informed by John Cardy that by perturbing the conformal
boundary condition it is possible to show the emergence of a  
mode dependent temperature directly in CFT.


\section{The infinite time limit and the generalized Gibbs ensemble} 
\label{sec:GGE}

In this section we consider the infinite time limit of the reduced density
matrix $\rho_A$ of a subsystem A composed of $\ell$ contiguous spins. 
%
To analyze $\rho_A$, we first consider its building
blocks, i.e. the two-point real-space correlation functions of fermions.
It is convenient to replace the Jordan-Wigner fermions $c_j$ (as
defined in Appendix A of paper I) by the (real) Majorana fermions 
\be
a_j^x=c_j^\dag+c_j\qquad a_j^y=i(c_j^\dag-c_j)\, ,
\label{majo}
\ee
which satisfy the algebra $\{a_l^x,a_n^x\}=2\delta_{l n}$, $\{a_l^y,a_n^y\}=2\delta_{l n}$, $\{a_l^x,a_n^y\}=0$.
In terms of these Majorana fermions, the operator $\sigma^x_j$ has the nonlocal representation 
\be
\sigma^x_\ell=\prod_{j=1}^{\ell-1}(i a_j^y a_j^x)a_\ell^x\, ,
\label{eq:sigmax}
\ee
while $\sigma_l^z$ is local
\be\label{eq:sigmaz}
\sigma_l^z=i a_l^y a_l^x\, .
\ee
The \emph{correlation matrix} $\Gamma$ is defined as
\be\label{eq:Gamma0}
\Gamma = \left[
 \begin{array}{ccccc}
\mathtt\Gamma_{0}  & \mathtt\Gamma_{-1}   &   \cdots & \mathtt\Gamma_{1-\ell}  \\
\mathtt\Gamma_{1} & \mathtt\Gamma_{0}   & &\vdots\\
\vdots&  & \ddots&\vdots  \\
\mathtt\Gamma_{\ell-1}& \cdots  & \cdots  &\mathtt\Gamma_{0}
\end{array}
\right], ~~~ 
\mathtt\Gamma_{l}=\left(
\begin{array}{cc}
-f_{l}&g_{l}\\
-g_{-l}&f_{l}
\end{array}
\right)\,,
\ee
where the matrix elements are the fermionic correlators
\begin{eqnarray}
g_n&\equiv& -\braket{a_l^x a_{l+n}^y},\nonumber\\
f_n+\delta_{n 0}&\equiv&\braket{a_{l}^x a_{l+n}^x}=\braket{a_{l}^y a_{l}^y}\qquad \forall l\, .
\label{corrferm}
\end{eqnarray}
The matrix $\Gamma$ is a block Toeplitz matrix, because its
constituent $2\times 2$ blocks depend only on the difference between
row and column indices. It is customary to introduce the (block)
\emph{symbol} of the matrix $\Gamma$ as follows 
\be
\label{eq:Gamma}
\hat\mathtt\Gamma(k)=
\left(
\begin{array}{cc}
-f(k)&g(k)\\
-g(-k)&f(k)
\end{array}
\right)\equiv
\sum_{n=-\infty}^\infty e^{-ink}\mathtt\Gamma_{n}.
\ee
The functions $f(k)$ and $g(k)$ are 
\begin{eqnarray}\label{eq:fg}
f(k)&=\sin\Delta_k \sin(2\veps_h(k) t),\nn
g(k)&=-i e^{i \theta_k}\Bigl[\cos\Delta_k-i\sin\Delta_k \cos(2\veps_h(k) t)\Bigr]\, ,
\end{eqnarray}
where $e^{i\theta_k}$ and $\cos\Delta_k$ are defined in \fr{Bangle} and
\fr{cosDeltak} respectively, while
\bea
\sin\Delta_k&=&\frac{(h-h_0)\sin k}{\sqrt{1+h^2-2 h \cos
    k}\sqrt{1+h_0^2-2 h_0 \cos k}}\, . 
\eea
The infinite time limit of the fermion correlators \fr{corrferm} is
straightforwardly obtained by Fourier transforming the functions
$f(k)$ and $g(k)$
\bea
f_j^\infty&=&  \braket{a_l^{x} a_{l+j}^{x}}\bigg|_{t=\infty}
-\delta_{j0}=0,\nn
\label{eq:gquench}
g_j^{\infty}&=& - \braket{a_l^x a_{l+j}^y}\bigg|_{t=\infty}
=-i\int_{-\pi}^\pi\frac{\mathrm d k}{2\pi}e^{-i j k}e^{-i\theta_k}\cos\Delta_k\, .
\eea
We can now use the Wick theorem to construct all correlation functions
in the Ising chain\footnote{For $t\rightarrow\infty$, the
expectation values of odd (with respect to fermion parity) operators
vanish for any quench. Thus all nonzero correlation functions can be
written in terms of two-point fermion correlators only.}.
Moreover, as shown in Refs
\cite{vidal,pe-rev}, the matrix $\Gamma$ determines the full
reduced density matrix of the block $A$ of $\ell$ contiguous fermions
(and hence spins, because contiguous spins are mapped to contiguous
fermions, see Eq. (\ref{eq:sigmax})) in the chain 
\be
\rho_A=\frac{1}{2^\ell} \sum_{\mu_l=0,1}
\Bigl< \prod_{l=1}^{2\ell} a_l^{\mu_l}\Bigr>\left(\prod_{l=1}^{2\ell}
a_l^{\mu_l}\right)^\dag \propto e^{a_l W_{lm} a_m/4}\,, 
\label{quad}
\ee
where $a_{2n}= a_{n}^x$, $a_{2n-1}=a_n^y$ and
\be
\tanh\frac{W}2=\Gamma\, .
\ee
Given $\rho_A$ one can calculate any local correlation function with
support in $A$. 

In order to understand the properties of the stationary reduced
density matrix, we should study the structure of the fermion
correlations in Eq. (\ref{eq:gquench}). These resemble the analogous
correlations of an Ising chain in equilibrium at finite temperature
$\beta^{-1}$ calculated in Ref. \cite{mc}
\bea\label{eq:gT}
f_j^{(\beta)}&=&  \langle\!\langle{a_l^x a_{l+j}^x}\rangle\!\rangle
-\delta_{j0}=0,\nn
g_j^{\beta}&=& - \langle\!\langle{a_l^x a_{l+j}^y}\rangle\!\rangle
=-i\int_{-\pi}^\pi\frac{\mathrm d k}{2\pi}e^{-i j
  k}e^{-i\theta_k}\tanh\Bigl(\frac{\beta \veps_k}{2}\Bigr)\, ,
\eea
where $\langle\!\langle{\cal O}\rangle\!\rangle$ denotes the
equilibrium expectation value of the operator ${\cal O}$ at
temperature $1/\beta$. Comparing \fr{eq:gT} to \fr{eq:gquench}
suggests that if it were possible to find a $\beta$ such that
$\cos\Delta_k=\tanh\bigl(\beta\veps_k/2\bigr)$ for any $k$,  
local properties of the (sub)system could be described by a
\emph{thermal state} at inverse temperature $\beta^{-1}$. It is easy
to see that this is not possible: on the one hand, the dispersion
relation is a continuous bounded function such that
$\tanh\bigl(\beta\veps_k/2\bigr)<1$ for any finite $\beta$.
On the other hand we have $|\cos\Delta_0|=|\cos\Delta_\pi|=1$, which implies
that the equality cannot be fulfilled at least in some neighbourhood
of the momenta $0$ and $\pi$.

Having ruled out the possibility of a thermal reduced density matrix
in the stationary state, we next investigate the conjecture put
forward by Rigol et. al. \cite{gg}, according to which the stationary
state of integrable models are described by a \emph{generalized Gibbs
ensemble}. The latter is obtained by maximizing the entropy, while
keeping the energy as well as all higher conservation laws fixed.
The resulting stationary reduced density matrix is of the form
\be
\rho_{\rm GGE}=\frac{1}{{\rm Z}_{\rm GGE}} e^{-\sum_m \lambda_m I_m}\,, 
\label{ggedef}
\ee
where the $I_m$ represent a complete set of independent, local
integrals of motion and the Lagrange multipliers $\lambda_m$ are fully
determined by the initial state $|\Psi_0\rangle$ through the conditions
\be
\Tr [\rho_{\rm GGE} I_m]=\langle \Psi_0|I_m | \Psi_0\rangle\,.
\ee
We stress that locality of $I_m$ is an essential feature of this
formulation. For the case of the Ising model, it is shown in
Ref.~\cite{EF12} that the local integrals of motion can be expressed as
\emph{linear} combinations of the Bogoliubov fermion number operators
$n_k\equiv \alpha_k^\dag\alpha_k$. Hence for the Ising model (and by
analogy other theories with free fermionic spectra) one may replace
the local charges $I_m$ in \fr{ggedef} by the number operators $n_k$,
despite the fact that the latter are non-local as discussed in the
introduction. Thus, taking the Lagrange multipliers as
$\lambda_k=\beta_k\veps_k$ in order to facilitate the eventual
identification of $\beta_k$ as a mode-dependent temperature, the GGE
takes the form
\be\label{eq:GGE2}
\rho_{\rm GGE}= \frac{1}{\rm Z_{\rm GGE}} {e^{-\sum_k \beta_k \veps_k
    \alpha_k^\dag\alpha_k}}\, . 
\ee
The Lagrange multipliers $\beta_k$ are fixed by the conditions
\be
\Tr [\rho_{\rm GGE}\ n_k]=\langle \Psi_0|n_k | \Psi_0\rangle\,,
\ee
which can be solved in closed form by
\be
\beta_k\veps_k=2\mathrm{arctanh}(\cos\Delta_k)\, .
\label{betak}
\ee
The fermionic two-point correlators in the GGE can be
straightforwardly evaluated following the finite-temperature
calculation and are given by
\be\label{eq:gGGE}
g_j^{(GGE)}=-i\int_{-\pi}^\pi\frac{\mathrm d k}{2\pi}e^{-i j k}e^{-i\theta_k}\tanh\Bigl(\frac{\beta_k \veps_k}{2}\Bigr)\, .
\ee
The crucial point now is that the fermionic two-point functions
$g_j^{\infty}$ and $g_j^{(GGE)}$ are exactly the same, given that the
Lagrange multipliers $\beta_k$ are related to $\Delta_k$ by
\fr{betak}, i.e.
\be
g_j^{\infty}=g_j^{(GGE)}.
\label{equality}
\ee
As the fermion correlators $f_j^\infty$ and $g_j^\infty$ completely
fix the reduced density matrix $\rho_A$ (the matrix $W_{lm}$ in
\fr{quad} can be determined by simply evaluating all fermionic two-point
functions using Wick's theorem), the equality \fr{equality} is \emph{lifted
to the level of the reduced density matrices}
\be
\rho_A={\rm Tr}_{\bar{A}}\big(\rho_{\rm GGE}\big) .
\ee
A characteristic feature of the GGE is that it retains detailed
information about the initial state. This is in contrast to a thermal
ensemble, which would correspond to $\beta_k=\beta$ for any $k$ and
would depend on the initial state only through the average value of
the energy. Given the structure of $\Delta_k$ there is no quench for
which the $\beta_k$ is independent on $k$ and hence the stationary
state is never thermal. However, it is possible for the differences
between GGE and thermal ensemble to be small for certain
observables \cite{rsms-08}. 

To summarize the results of this section, we have shown that the
stationary reduced density matrix after an arbitrary quench of the
bulk magnetic field in the Ising chain is equivalent to a generalized Gibbs
ensemble. Hence all local correlation functions can be deduced from
the GGE. The calculation of observables in the framework of the GGE
generically remains a difficult problem. In the following sections we
determine the stationary behaviour of some of the most important
observables, namely the longitudinal and transverse correlation
functions.

\section{Stationary value of the longitudinal correlation function}  %
\label{sec:long}

The two-point function of $\sigma^x_j$ is the expectation value of a
string of Majorana fermions 
\be
\rho^{xx}(\ell,t)=\braket{\prod_{j=1}^\ell (-i a_j^y(t) a_{j+1}^x(t))}\, ,
\label{rhoxxF}
\ee
and by 
Wick's theorem one obtains a representation as the Pfaffian of the
$2\ell\times 2\ell$ antisymmetric matrix  $\bar\Gamma$ considered in Ref.~\cite{mc} 
\be\label{eq:rhoxxPf}
\rho^{xx}(\ell,t)=\mathrm{pf}(\bar\Gamma)\, ,
\ee
which has the same structure of the correlation matrix in Eq. (\ref{eq:Gamma0})  with blocks
\be
\bar\Gamma_l= \left(
\begin{array}{cc}
-i f_{l}&-ig_{l-1}\\
i g_{-l-1}&i f_{l}
\end{array}
\right)\, .
\ee
From the asymptotic correlations (\ref{eq:gquench}), we can compute
this correlation function in the limit of infinite time. Since
$f_n^\infty=0$, the block structure in the matrix $\bar\Gamma$ simplifies,
and the two-point function of the order parameter in the infinite time
limit becomes the determinant of a $\ell\times\ell$ Toeplitz matrix
$G$ 
\be
\rho^{xx}(\ell,t)=\det\big(G\big)\ ,\quad
G_{l n}=g^\infty_{l-n}\ .
\label{matrixG}
\ee
The symbol of $G$ is
\be
g_\infty(k)=-e^{i(k-\theta_k)}\cos\Delta_k\, .
\ee
The representation \fr{matrixG} is more convenient than
\fr{eq:rhoxxPf} for determining the large-$\ell$ asymptotics of
$\rho^{xx}(\ell,t)=\det\big(G\big)$. We will show that at late times
the two-point function of the order parameter decays exponentially
with the distance  
\be
\rho^{xx}(\ell\gg1,t=\infty)= C^x(\ell)
e^{-\ell/\xi}\left[1+o\big(\ell^0\big)\right]. 
\label{rhoexp2}
\ee
%
The result \fr{rhoexp2} is obtained by using exact results on the
asymptotic behaviour of the determinant of Toeplitz matrices, such as
Sz\"ego's lemma, the Fisher-Hartwig conjecture, and generalizations
thereof (the results we need in the following are summarized
in \ref{szapp} and are taken from Refs. \cite{FH-gen,FH-DIK,Ehr,sviaj}). 
To apply these theorems it is useful to change variable as $z=e^{i
  k}$, so that the symbol can be rewritten as  
\be\label{eq:symbolinf}
\fl \qquad \mathfrak g_\infty(z)\equiv g_\infty(-i\ln z)=
\frac{h+h_0}{2\sqrt{h_0}}\frac{\sqrt{z}}{\sqrt{h_0-z}\sqrt{z-1/h_0}}\frac{(z-h_1)(1/h_1-z)}{z-h}\, ,
\ee
where $h_1$ is defined in \Eref{eq:h1}.
The  asymptotic behaviour of the determinant and so the  physical quantities 
$\xi$ and $C^x(\ell)$ depend on the analytic properties of the symbol, 
which in turn change with the quench parameters.

\subsection{Quenches within the ferromagnetic phase.}

\begin{figure}[t]
\begin{center}
\includegraphics[width=0.7\textwidth]{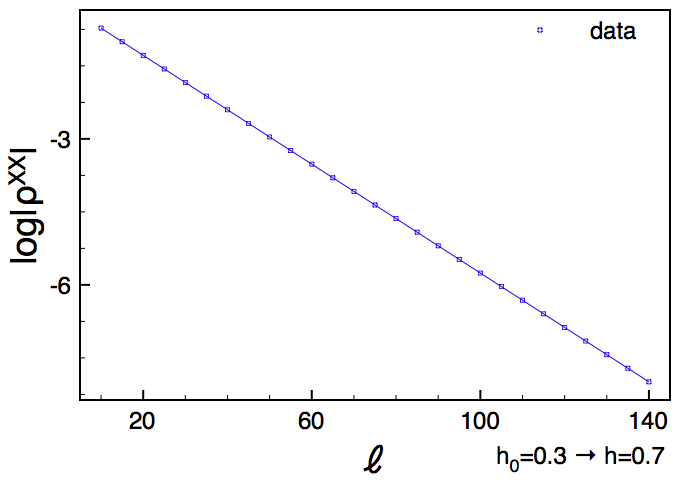}
\caption{$\ln |\rho^{xx}(\ell, t=\infty)|$ after a quench within the
ordered phase from $h_0=0.3$ to $h=0.7$.  
The straight line is the asymptotic prediction in Eq. (\ref{rhoasyFF}). }
\label{fig:xi_O} 
\end{center}
\end{figure}

The symbol (\ref{eq:symbolinf}) is different from zero on the unit circle $|z|=1$ and has winding number zero about 
the origin $z=0$.
Under these conditions,  it is possible to apply the strong Sz\"ego
lemma (see (\ref{eq:Szego})).
The inverse correlation length is straightforwardly obtained as
\be
\xi^{-1}=-\int_{0}^{2\pi}\frac{\mathrm d k}{2\pi}\ln  \mathfrak g_\infty(e^{i k}) =-\int_0^{2\pi}\frac{\mathrm d k}{2\pi}\ln\cos\Delta_k\, .
\ee
The integral can be carried out using ($x>0$)
\be
\int_{0}^{2\pi}\frac{\mathrm d k}{2\pi}\ln(x-e^{i k})=\theta(x-1)\ln x\, ,
\ee
to finally obtain
\be\label{eq:xi_O}
\xi^{-1}=-\ln\Bigl(\frac{h+h_0}{2}h_1\Bigr)\, .
\ee
The multiplicative factor $C^x(\ell)={\cal C}^x_{\rm FF}$ is slightly
more complicated to obtain. It is given by (see Eq. (\ref{eq:Szego})) 
\be
{\cal C}^x_{\rm FF}= \exp\Bigl[\sum_{k\geq 1}k (\ln  \mathfrak
  g_\infty)_k(\ln  \mathfrak g_\infty)_{-k}\Bigr]\, , 
\label{FF1}
\ee
where we used the same notation as in the appendix. 
The Fourier coefficients $(\ln \mathfrak g_\infty)_k$ are 
\be
 (\ln  \mathfrak g_\infty)_k=\left\{
\begin{array}{ll}
(h_0^k-2 h_1^{-k})/ (2 k) &{\rm if\ \ }k>0\ ,\\
\ln\bigl((h+h_0)h_1/2\bigr)&{\rm if\ \ }k=0\ ,\\
(2 h_1^{k}-2 h^{-k}-h_0^{-k})/(2k)&{\rm if\ \ }k<0\, .
\end{array}\right.
\ee
Carrying out the $k$-sum in \fr{FF1} we obtain
\be
{\cal C}^x_{\rm FF}=\frac{1-h h_0+\sqrt{(1-h^2)(1-h_0^2)}}{2\sqrt{1-h
    h_0}\sqrt[4]{1-h_0^2}}. 
\ee
Putting everything together we conclude that the two-point function
has the following asymptotic behaviour 
\be
\rho^{xx}(\ell,t=\infty)= \frac{1-h h_0+\sqrt{(1-h^2)(1-h_0^2)}}{2\sqrt{1-h h_0}\sqrt[4]{1-h_0^2}}e^{-\ell/\xi}+O(e^{-\kappa \ell})\,,
\label{rhoasyFF}
\ee
for some $\kappa>\xi^{-1}$.
In Fig. \ref{fig:xi_O} this analytic prediction is compared with
numerical data. The agreement is clearly excellent.

\subsection{Quenches within the paramagnetic phase.}
Here the winding number of the symbol equals $1$.
Following Ref.~\cite{sviaj} (see \ref{szapp} for some details) we first
rewrite the symbol by factorizing an appropriate power of $z$, such
that the remainder has winding number zero about the origin
\be
y(z)\equiv
-\mathfrak g_\infty(1/z) z=\frac{h+h_0}{2h\sqrt{h_0}}\frac{\sqrt{z}}
{\sqrt{z-1/h_0}\sqrt{h_0-z}}\frac{(z-1/h_1)(h_1-z)}{z-1/h}\, . 
\ee
The function $y(z)$ has winding number zero as desired and no zeroes on
the unit circle $|z|=1$. We note that $\mathfrak g_\infty(1/z)$ is in
fact the symbol of the transpose of $G$. It has negative winding
number, which is precisely the case discussed in \ref{szapp}.
From the Wiener-Hopf factorization of $y(z)$ as defined in Eq. (\ref{WHf})
\bea
\label{eq:y-y+}
y(z)&= y_+(z) \frac{h+h_0}{2 h h_0}h_1\, y_-(z)\, ,\nn
y_\pm (z)&=\exp\sum_{k\geq 1}(\ln  y)_{\pm k}z^{\pm k},
\eea
the correlation function is obtained using (\ref{FHwnm1})
\be\label{eq:PP}
\rho^{xx}(\ell,t=\infty) \simeq \exp\Bigl[\sum_{k\geq 1}k (\ln  y)_k(\ln  y)_{-k}\Bigr]\Bigl(\frac{h+h_0}{2 h h_0}h_1\Bigr)^{\ell}  I_\ell \, .
\ee
Here $I_\ell$ is given in terms of $y_\pm(z)$ by
\be
I_\ell= \int_0^{2\pi} \frac{\mathrm d k}{2\pi }\ e^{i \ell k}  
\frac{y_-\big(e^{-i k}\big)}{y_+\big(e^{-i k}\big)}\,.
\label{Iell}
\ee
The inverse correlation length can be read off from \fr{eq:PP} to be
\be\label{eq:xidis}
\fl\qquad\qquad\xi^{-1}=-\ln \Bigl(\frac{h+h_0}{2 h h_0}h_1\Bigr)-\lim_{\ell\rightarrow\infty}\frac{1}{\ell}\ln I_\ell\, .
\ee
In order to find $y_-(z)$ we use the standard Wiener-Hopf factorizations
\be\label{eq:factorizations}
(\alpha (x- z)^q)_-=\left\{\begin{array}{ll}
\bigl(\frac{z-x}{z-1}\bigr)^q&{\rm if\ \ }x<1\ ,\\
1&{\rm if\ \ }x>1\ ,
\end{array}\right.
\ee
\be
(\alpha(x-z)^q)_+=\left\{\begin{array}{ll}
(1-z)^q&{\rm if\ \ }x<1\ ,\\
(1-z/x)^q&{\rm if\ \ }x>1\, , 
\end{array}\right.
\ee
where  $\alpha$ is an arbitrary complex number. 
The rhs above are independent on $\alpha$ because of our normalization of the factorization in Eq. (\ref{WHf}).

The Wiener-Hopf
factorization of $ y(z)$ is a combination of the above with
$q=0,\pm{1}/{2},\pm 1$, and after some simple algebra we get
\be
y_-(z)=\frac{\sqrt{z}}{\sqrt{z-1/h_0}}\frac{z-1/h_1}{z-1/h},\qquad y_+(z)=\frac{\sqrt{h_0}}{h_1}\frac{h_1-z}{\sqrt{h_0-z}}\, .
\ee
Substituting these expressions into $I_\ell$, we obtain
\be\label{eq:I}
I_\ell=\frac{h\sqrt{h_0}}{h_1}\oint_{C}\frac{\mathrm d z}{2\pi i} z^{\ell-\frac{1}{2}}\frac{\sqrt{z-1/h_0}}{\sqrt{h_0-z}}\frac{h_1-z}{(h-z)(z-1/h_1)}\, ,
\ee
where $C$ is the unit circle. For numerical computations the following
equivalent expression is more convenient
\be
I_\ell=2 J h\int_0^{2\pi}\frac{\mathrm d k}{2\pi}\frac{e^{i\ell k}}{\varepsilon_h(k)}e^{-i (\theta(k)+\theta_0(k)-2 \theta_1(k))}\, ,
\ee
where $\theta_1$ is the Bogoliubov angle corresponding to the magnetic
field $h_1$, i.e. 
\be
e^{i\theta_1(k)}=\frac{h_1-e^{i k}}{\sqrt{1+h_1^2-2 h_1\cos k}}\, .
\ee
The integral \fr{eq:I} is dominated by the vicinities of the points in
the interior of the unit circle, at which the integrand is non-analytic. 
There are two branch points at $z=0$ and $z=1/h_0$, and a simple pole
at $z=1/h_1$. If the leading contribution arises from $z\approx 1/h_1$, the
$\ell$-dependence takes a simple exponential form, while additional
power-law corrections arise if the integral is dominated by the
region $z\approx 1/h_0$. The leading exponential contribution arises
from the pole if $h_0>h_1$ (corresponding to $h_0>h$) and by the
branch point at $z=1/h_0$ if $h_0<h_1$ (corresponding to $h>h_0$)
\be
I_\ell \propto \left\{\begin{array}{lr}
h_1^{-\ell}& h_0>h\ ,\\
 h_0^{-\ell}&h_0<h\, .
 \end{array}\right.
\ee
Combining this result with \fr{Iell} and \Eref{eq:xidis} leads to the
following result for the inverse correlation length
\be\label{eq:xiD}
\xi^{-1}=-\ln\Bigl(\frac{h+h_0}{2 h h_0}h_1\Bigr)+\ln\min(h_0,h_1)\, .
\ee

\subsubsection{Calculation of the constant ${\cal C}^x_{\rm PP}$ for
  $h_0>h$.} 
From (\ref{eq:PP}) we have
\be
{\cal C}^x_{\rm PP}=B \exp\Bigl[\sum_{k\geq 1}k (\ln y)_k(\ln
   y)_{-k}\Bigr] \, , 
\label{CPP1}
\ee
where $B$ is the residue of the integrand in \Eref{eq:I}
at $1/h_1$ without the exponential factor $h_1^{-\ell}$
\be 
 B =\frac{2h \sqrt{h_0^2-1}}{(h_0-h)\sqrt{h_0^2-h^2}}\Bigl(h
 h_0-1-\sqrt{(h^2-1)(h_0^2-1)}\Bigr)\ .
\ee
The Fourier coefficients $(\ln  y)_{k}$ are the same as for the quench
from $1/h_0$ to $1/h$, because $\cos\Delta_k$ is invariant under the
inversion of the magnetic fields, so that
\be
(\ln  y)_{k}\bigg|_{h_0\rightarrow h}=
(\ln \mathfrak g_{\infty}^{\rm FF})_{k}\bigg|_{1/h_0\rightarrow 1/h}\ ,
\qquad k\neq 0\, .
\ee
Using this in \fr{CPP1} we conclude that the ``Sz\"ego part'' of
${\cal C}^x_{\rm PP}$ is the same as in the ordered phase 
\be
\exp\Bigl[\sum_{k\geq 1}k (\ln  y)_k(\ln  y)_{-k}\Bigr]=
{\cal C}_{\rm FF}(1/h_0,1/h)\, .
\ee
Combining the two pieces in \Eref{CPP1} we have
\be
{\cal C}^x_{\rm PP}=
\sqrt{\frac{h(h_0-h)\sqrt{h_0^2-1}}{(h+h_0)(h h_0-1)}}\, ,\qquad h_0>h,
\ee 
so that the two-point function has the following asymptotic behaviour
\be
\rho^{xx}(\ell,t=\infty)=\sqrt{\frac{h(h_0-h)\sqrt{h_0^2-1}}{(h+h_0)(h h_0-1)}}e^{-\ell/\xi}+O(e^{-\tilde{\kappa} \ell}),\qquad h_0>h,
\label{eq:h0greath}
\ee
where $\tilde{\kappa}>\xi^{-1}$.
\begin{figure}[t]
\begin{center}
\includegraphics[width=0.48\textwidth]{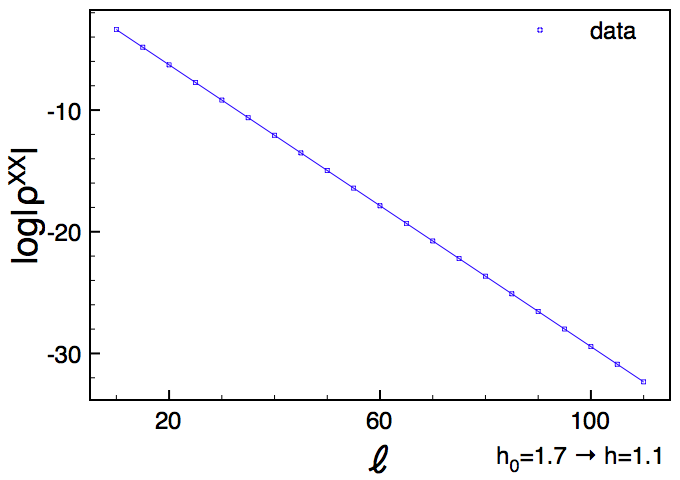}
\includegraphics[width=0.48\textwidth]{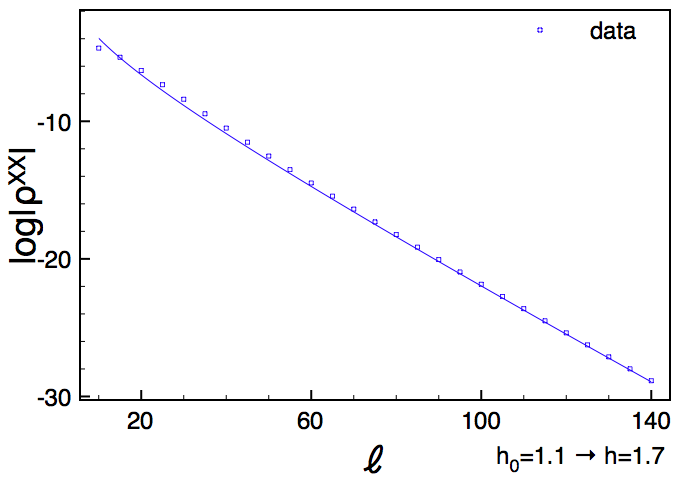}
\caption{$\ln |\rho^{xx}(\ell, t=\infty)|$ as function of $\ell$ 
after two quenches within the disordered phase. 
Left: The case $h_0>h$: The solid line is the prediction
(\ref{eq:h0greath}) and is seen to be in excellent agreement with the
numerical data.
Right: The case $h_0<h$. The straight line is the leading term of the prediction (\ref{eq:h0lessh}). 
Some deviations at small $\ell$ are visible due to the subleading correction in (\ref{eq:h0lessh}).
}\label{fig:xi_D} 
\end{center}
\end{figure}
The asymptotic result \fr{eq:h0greath} is compared to numerical
results for the two-point function in Fig.~\ref{fig:xi_D} (left
panel). The agreement is clearly excellent.

\subsubsection{Calculation of ${\cal C}^x_{\rm PP}(\ell)$ for $h_0<h$.}
This case is slightly more involved. The two point function acquires a
power law prefactor since the leading contribution to the integral
(\ref{eq:I}) arises from the vicinity of the branch point at $1/h_0$.
Isolating this contribution gives
\be
\fl\quad I_\ell\approx -\frac{2 h}{\, h_1}h_0^{-\frac{1}{2}-\ell}\int_{\frac{h_0}{h_1}+\epsilon_1+i\epsilon}^{1+i\epsilon}\frac{\mathrm d x}{2\pi i}\frac{x^{\ell-\frac{1}{2}} \sqrt{x-1}}{\sqrt{h_0-x/h_0}}\frac{h_1-x/h_0}{(h-x/h_0)(x/h_0-1/h_1)}+O(h_1^{-\ell})\, ,
\ee
where $\epsilon$ and $\epsilon_1$ are two infinitesimal positive
constants. After the change of variable $w=(1-x)\ell_s$, with
$\ell_s=\ell-\frac{1}{2}$, we obtain
\be\label{eq:series}
\fl\quad I_\ell\approx -\frac{2 h }{\, h_1}\frac{h_0^{-1-\ell_s}}{\ell_s^{\frac{3}{2}}}\int_0^{M_\ell}\frac{\mathrm d w}{2\pi}\frac{\sqrt{w} (1-\frac{w}{\ell_s})^{\ell_s}}{\sqrt{h_0-\frac{1}{h_0}+\frac{w}{\ell_s h_0}}}\frac{h_1-\frac{1}{h_0}+\frac{w}{\ell_s h_0}}{(h-\frac{1}{h_0}+\frac{w}{\ell_s h_0})(\frac{1}{h_0}-\frac{1}{h_1}-\frac{w}{\ell_s h_0})},
\ee
where
\be
M_\ell=\ell_s\Bigl(1-\frac{h_0}{h_1}-\epsilon_1\Bigr)\, .
\ee
The large $\ell$ asymptotics is obtained by expanding in powers of
$1/\ell_s$ and taking the integration limit ($M_\ell$) to infinity. 
At a given order of the asymptotic expansion the resulting error is
exponentially small. After straightforward algebra we arrive at
\be
I_\ell\approx -\frac{\Gamma(3/2)}{2\pi} \frac{2 h h_0}{\sqrt{h_0^2-1}}\frac{h_0 h_1-1}{(h_0 h-1)(h_1-h_0)}\frac{h_0^{-\ell}}{(\ell-\frac{1}{2})^{\frac{3}{2}}},
\ee
which in turn leads to the following result for the two-point function
\bea
\label{eq:h0lessh}
\fl\qquad
\rho^{xx}(\ell,t=\infty)\simeq& -
\frac{h_0\sqrt{ h}\Bigl(h h_0-1+\sqrt{(h^2-1)(h_0^2-1)}\Bigr)^2}{4
  \sqrt{\pi } (h_0^2-1)^{3/4} (h_0 h-1)^{3/2} (h-h_0)}\nn
&\times
\bigl(\ell-\frac{1}{2}\bigr)^{-\frac{3}{2}}
\Bigl(1+\frac{\alpha}{\ell-\frac{1}{2}}+\dots\Bigr)e^{-\ell/\xi},\quad h_0<h\, .
\eea
The constant $\alpha$ characterizing the subleading contribution in
\fr{eq:h0lessh} is given by
\be
\fl\qquad \alpha=
\frac{3}{8}\Bigl[\frac{3}{h_0^2-1}-\frac{4}{h_0 h-1}-\frac{h_0(5h_0(h-h_0)-8\sqrt{(h^2-1)(h_0^2-1)})}{(h-h_0)(h_0^2-1)}\Bigr]\, .
\ee
The requirement that the term involving $\alpha$ should be small
establishes the regime of validity of the expansion \fr{eq:h0lessh}.
In particular, for small quenches in the sense of paper I,
i.e. $\eta=\max|\sin\Delta_k|$ being small, we have
\be
\alpha=\frac{3 h}{2(h^2-1)}\frac{1}{\eta}+O(\eta^0)\ .
\ee
Hence \fr{eq:h0lessh} holds only for distances that are sufficiently
large compared to $(h-h_0)^{-1}$. In particular, in the limit $h\to
h_0$ its region of validity disappears entirely.

In Fig.~\ref{fig:xi_D} the asymptotic result \fr{eq:h0lessh} (with $\alpha$ set to zero) is
compared to numerical data. It is evident that for $h_0<h$ and
$\ell$ not large enough there are sizeable corrections to 
(\ref{eq:h0lessh}).

\subsection{Quenches across the critical point.}
\begin{figure}[t]
\begin{center}
\includegraphics[width=0.7\textwidth]{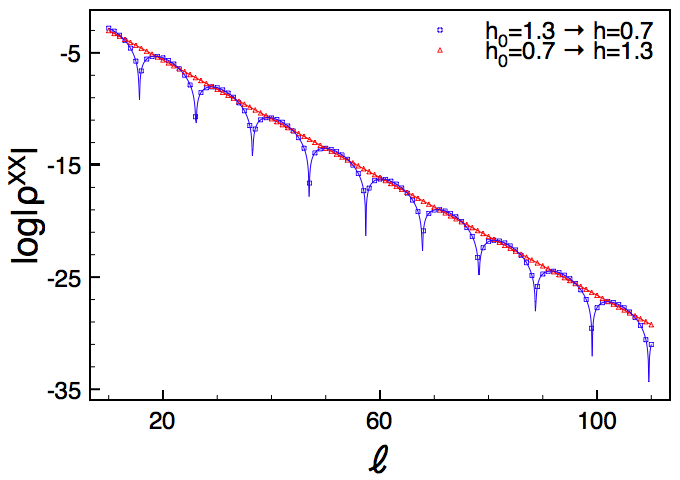}
\caption{$\ln |\rho^{xx}(\ell, t=\infty)|$ as a function of $\ell$ 
for quenches across the critical point. Blue squares correspond to a
quench from the paramagnetic ($h_0=1.3$) to the ferromagnetic
($h=0.7$) phase. The blue solid line is the asymptotic expression
(\ref{eq:quenchPF}). Red triangles correspond to a quench from the
ferromagnetic ($h_0=0.7$) to the paramagnetic ($h=1.3$) phase. The red
straight line is the asymptotic result (\ref{eq:xido}). In both cases
the asymptotic and numerical results are seen to be in excellent
agreement.
}\label{fig:xi_across} 
\end{center}
\end{figure}

For quenches across the critical point $h_1$ is a complex number with unit modulus $|h_1|=1$. 
In this case, the symbol has two simple zeroes on the integration contour  at $z_1=h_1$ and $z_2=1/h_1=h_1^*$.  
Thus, as first step we need to cast the symbol in the form (\ref{eq:symbolform1}) which is suitable for application
of the Fisher-Hartwig conjecture as reported in (\ref{eq:FH}). 
Since the symbol has no discontinuities, we only need to isolate the zeroes.
This can be done by rewriting $g_\infty(z)$ in the form
\be\label{eq:FHform}
\fl\qquad
\mathfrak{g}_\infty(z)=\left[h_1^{-\beta_1}(h_1^\ast)^{-\beta_2}\right]
f^{(\beta_1,\beta_2)}(z)\
z^{\beta_1+\beta_2}\ (-1)^{\beta_1-\beta_2}\
\frac{(h_1-z)(z-h_1^\ast)}{z}\ ,
\ee
where 
\be\label{eq:gw0}
\fl\qquad 
f^{(\beta_1,\beta_2)}(z)=\left[h_1^{\beta_1}(h_1^\ast)^{\beta_2}
\frac{h+h_0}{2\sqrt{h_0}}\right]
(-1)^{\beta_1-\beta_2}\
\frac{\sqrt{z}}{\sqrt{h_0-z}\sqrt{z-h_0^{-1}}}\frac{z^{1-\beta_1-\beta_2}}{z-h}\ .
\ee
The symbol \fr{eq:FHform} is now in the desired form \fr{eq:symbolform1}
with
\be\fl\qquad
R=2\ ,\quad
\alpha_0=\beta_0=0\ , \quad
\alpha_1=\frac{1}{2}\ ,\quad
\beta_{1}=\frac{1}{2}-m_{1}\, ,\quad
\alpha_2=\frac{1}{2}\ ,\quad
\beta_{2}=\frac{1}{2}-m_{2}\, ,
\label{eq:beta12}
\ee
where $m_{1,2}\in\mathbb{Z}$ are integers that label
the inequivalent representations of the symbol.
Taking into account the contributions of all representations
results in an expression for the two-point function of the form
\be\label{eq:rhox}
\fl\qquad\rho^{xx}(\ell\gg1,t=\infty)\simeq 
\sum_{\{\beta_1,\beta_2\}_{\mathrm{ineq.}}}
C_{\beta_1,\beta_2}\ \ell^{1/2-\beta_1^2-\beta_2^2}\
\left[-h_1^{\beta_1}(h_1^\ast)^{\beta_2}\right]^{\ell}\
e^{- \ell/\xi(\beta_1,\beta_2)}\, ,
\ee
where we isolated the oscillatory factor 
$\big[-h_1^{\beta_1}(h_1^\ast)^{\beta_2)}\big]^{\ell}$. In \fr{eq:rhox},
$\xi(\beta_1,\beta_2)$ and $C_{\beta_1,\beta_2}$ are respectively
the correlation length and correlation amplitude corresponding to the
representation labelled by $\beta_{1,2}$ of the symbol (\emph{cf}.  
(\ref{eq:FH})). In the special case $R=2$, $\alpha_1=\alpha_2={1}/{2}$, 
$\alpha_0=\beta_0=0$, $z_1=h_1$, $z_2=h_1^\ast$, the amplitudes of the
various representations are given by
\begin{eqnarray}\label{eq:constant}
\fl\quad
C_{\beta_1,\beta_2}=& \exp\Bigl[\sum_{q\geq 1}q \big(\ln
  f^{(\beta_1,\beta_2)}\big)_q \
\big(\ln  f^{(\beta_1,\beta_2)}\big)_{-q}\Bigr] 
\left[f^{(\beta_1,\beta_2)}_+(h_1)\right]^{-\frac{1}{2}+\beta_1}
\left[f^{(\beta_1,\beta_2)}_-(h_1)\right]^{-\frac{1}{2}-\beta_1}\nn
\fl\quad&\times
\left[f^{(\beta_1,\beta_2)}_+(h_1^\ast)\right]^{-\frac{1}{2}+\beta_2}
\left[f^{(\beta_1,\beta_2)}_-(h_1^\ast)\right]^{-\frac{1}{2}-\beta_2}
\nn\fl\quad
&\times
|h_1-h_1^\ast|^{2(\beta_1\beta_2-\frac{1}{4})}\Bigl(\frac{h_1^\ast}{h_1 e^{i\pi}}\Bigr)^{\frac{\beta_2}{2}-\frac{\beta_1}{2}}\prod_{j=0}^2\frac{G(\frac{3}{2}+\beta_j)G(\frac{3}{2}-\beta_j)}{G(2)}\, ,
\end{eqnarray}
where $G$ is Barnes' G-function and $f^{(\beta_1,\beta_2)}_\pm(z)$ are
the factors of the Wiener-Hopf factorization of
$f^{(\beta_1,\beta_2)}(z)$ in the conventions \fr{WHf}. As $G(-n)=0$
for $n=0,1,2\dots$ we need to consider only the four inequivalent
representations 
\be
\beta_1=\pm\frac{1}{2}\ ,\quad
\beta_2=\pm\frac{1}{2}\ .
\ee
In order to obtain explicit expressions for
$C_{\pm\frac{1}{2},\pm\frac{1}{2}}$ and
$\xi({\pm\frac{1}{2},\pm\frac{1}{2}})$ we need to consider quenches
from the paramagnetic to the ferromagnetic phase and vice versa
separately. This is done in the following two subsections.

\subsubsection{Quenches from the paramagnetic to the ferromagnetic
  phase.}
The leading contributions in the large-$\ell$ limit arise from the two
representations 
\be
(\beta_1,\beta_2)=\big(\frac{1}{2},-\frac{1}{2}\big)\ ,\qquad
(\beta_1,\beta_2)=\big(-\frac{1}{2},\frac{1}{2}\big)\ .
\ee
In both cases the inverse correlation length takes the simple form
\be\label{eq:xido}
\xi^{-1}\Bigl(\pm\frac{1}{2},\mp\frac{1}{2}\Bigr)=-\int_{0}^{2\pi}\frac{\mathrm d k}{2\pi}\ln \Bigl(h_1^{\mp 1}f^{(\pm\frac{1}{2},\mp \frac{1}{2})}(e^{i k})\Bigr)=-\ln\Bigl(\frac{h+h_0}{2 h_0}\Bigr)\, .
\ee
As shown below, the constants $C_{\beta_1,\beta_2}$ are equal to
\be
C_{\frac{1}{2},-\frac{1}{2}}=C_{-\frac{1}{2},\frac{1}{2}}
=\frac{1}{2}\sqrt{\frac{h_0-h}{\sqrt{h_0^2-1}}}
\, .
\label{E12}
\ee
Putting everything together we arrive at the following result for the
asymptotics of the two-point function
\be\label{eq:quenchPF}
\rho^{xx}(\ell,t=\infty)=
\sqrt{\frac{\ h_0-h}{\sqrt{h_0^2-1}}}\
\mathrm{Re}[h_1^\ell]\ e^{-\ell/\xi}+O(e^{-\kappa'\ell})\, ,
\ee
where $\kappa'>\xi^{-1}$. Rewriting $h_1$ as a function of $h$ and
$h_0$ one recovers the form given in (\ref{CPFcos}). The 
expression \fr{eq:quenchPF} is compared to numerical results for
$\rho^{xx}(\ell,t=\infty)$ in Fig. \ref{fig:xi_across}. The agreement
for large distances $\ell$ is clearly excellent. We note that
\fr{eq:quenchPF} can be understood as the leading term in an
asymptotic expansion only if $\mathrm{Re}[h_1^\ell]$ is of order 1. 
For the particular quench from
\be
h_0=\frac{h+\sqrt{2}(1-h^2)}{1-2 h^2}
\ee
to $h<1/\sqrt{2}$ we have $h_1^2=i$ and (\ref{eq:quenchPF}) vanishes
for $\ell=4 n+2$, with $n$ integer. For this particular sequence of
distances $\xi$ does then not represent the correlation length.

\paragraph{Calculation of the constant $C_{\frac{1}{2},-\frac{1}{2}}$.}
The Wiener-Hopf factorization of $f^{(\frac{1}{2},-\frac{1}{2})}(z)$ is
\be
f^{(\frac{1}{2},-\frac{1}{2})}(z)=
 f^{(\frac{1}{2},-\frac{1}{2})}_+(z)\frac{(h+h_0)h_1}{2 h_0}
f^{(\frac{1}{2},-\frac{1}{2})}_-(z)\, ,
\ee
where
\be
f^{(\frac{1}{2},-\frac{1}{2})}_+(z)=
\frac{\sqrt{h_0}}{\sqrt{h_0-z}}, \qquad 
f^{(\frac{1}{2},-\frac{1}{2})}_-(z)=
\frac{\sqrt{z}}{\sqrt{z-1/h_0}}\frac{z}{z-h}\, .
\label{WHFpm}
\ee
This results in Fourier coefficients
\be
\big(\ln f^{(\frac{1}{2},-\frac{1}{2})}\big)_k=
\left\{\begin{array}{ll}
h_0^{-k}/ (2 k)  &{\rm if\ \ }k>0\ ,\\
\ln\bigl(\frac{h+h_0}{2 h_0}h_1\bigr)&{\rm if\ \ }k=0\ ,\\
-(h_0^k+2 h^{-k})/(2 k)&{\rm if\ \ }k<0\, .
\end{array}\right.
\ee
The Sz\"ego part of the constant is then of the form
\be
 \exp\Bigl[\sum_{k\geq 1} k \big(\ln
   f^{(\frac{1}{2},-\frac{1}{2})}\big)_k 
\big(\ln
   f^{(\frac{1}{2},-\frac{1}{2})}\big)_{-k}\Bigr]=
 \frac{h_0}{\sqrt[4]{h_0^2-1}\sqrt{h_0-h}}\, . 
\label{WHFpm2}
\ee
Substituting \fr{WHFpm} and \fr{WHFpm2} into the expression
(\ref{eq:constant}) we obtain the result (\ref{E12}). The calculation
of $C_{-\frac{1}{2},\frac{1}{2}}$ is completely analogous.

\subsubsection{Quenches from the ordered to the disordered phase.} 
Here the leading contribution in (\ref{eq:rhox}) arises from
the representation 
\be
\beta_1=\beta_2=\frac{1}{2}.
\ee
The Wiener-Hopf factorization  of $f^{(\frac{1}{2},\frac{1}{2})}(z)$ is
\be
f^{(\frac{1}{2},\frac{1}{2})}(z)=
f^{(\frac{1}{2},\frac{1}{2})}_+(z)
\frac{h+h_0}{2 h}
f^{(\frac{1}{2},\frac{1}{2})}_-(z)\ ,
\ee
where
\be
f^{(\frac{1}{2},\frac{1}{2})}_+(z)=\frac{1}{\sqrt{1-z
    h_0}}\frac{1}{1-z/h},\qquad 
f^{(\frac{1}{2},\frac{1}{2})}_-(z)=\frac{\sqrt{z}}{\sqrt{z-h_0}}\, .
\label{whf3}
\ee
The Fourier coefficients of $\ln f^{(\frac{1}{2},\frac{1}{2})}$ are 
\be
\big(\ln f^{(\frac{1}{2},\frac{1}{2})}\big)_k
=\left\{\begin{array}{ll}
(h_0^k+2 h^{-k})/ (2 k) &{\rm if\ \ }k>0\ ,\\
\ln\bigl(\frac{h+h_0}{2 h}\bigr)&{\rm if\ \ }k=0\ ,\\
-h_0^{-k}/(2k)&{\rm if\ \ }k<0\, .
\end{array}\right.
\ee
while the inverse correlation length is 
\be
\xi^{-1}= \big(\ln f^{(\frac{1}{2},\frac{1}{2})}\big)_0
=\ln\Bigl(\frac{h+h_0}{2 h}\Bigr)\,.
\label{corrlFP}
\ee
We note that \fr{corrlFP} coincides with the result obtained by formally
interchanging $h_0$ with $h$ for quenches from the disordered to the
ordered phases (\ref{eq:xido}). The Sz\"ego part of the constant is
given by 
\be
 \exp\Bigl[\sum_{k\geq 1} k 
\big(\ln 
f^{(\frac{1}{2},\frac{1}{2})}
\big)_k \big(\ln f^{(\frac{1}{2},\frac{1}{2})}\big)_{-k}\Bigr]=
\frac{\sqrt{h}}{\sqrt[4]{1-h_0^2}\sqrt{h-h_0}}\, .
\label{whf4}
\ee
Substituting \fr{whf3} and \fr{whf4} into (\ref{eq:constant}) and
\fr{whf3} we obtain the following result for the asymptotic behaviour
of the two-point function 
\be
\label{eq:quenchFP}
\rho^{xx}(\ell,t=\infty)=\sqrt{\frac{h\sqrt{1-h_0^2}}{h+h_0}}
e^{-\ell/\xi}+O(e^{-\bar{\kappa}\ell})\, , 
\ee
where $\bar{\kappa}>\xi^{-1}$. In Fig. \ref{fig:xi_across} the analytic
result \fr{eq:quenchFP} is compared to a numerical computation
of $\rho^{xx}(\ell,t=\infty)$ and the agreement is seen to be excellent.

\subsection{A general formula for the correlation length.}
The results \fr{eq:xi_O}, \fr{eq:xiD}, \fr{eq:xido} and \fr{corrlFP} 
can be combined into the following general expression for the inverse
correlation length $\xi^{-1}$  
\be
\fl\qquad\xi^{-1}=-\int_0^{2\pi}\frac{\mathrm d k}{2\pi}\ln|\cos\Delta_k|+\theta_H(h-1)\theta_H(h_0-1)\ln\min(h_0,h_1)\, ,
\ee
where $h_1$ is defined in \Eref{eq:h1}. After simple algebraic
manipulations, this expression can be cast in the form reported in
(\ref{xiasy}). 
%

\section{Transverse correlations} %
\label{sec:tran}
We now turn to the correlation functions of transverse spins
$\sigma_l^z$. These are simple because $\sigma_l^z$ are \emph{local}
in the fermionic representation $\sigma_l^z=i a_l^y a_l^x$.
Concomitantly, and in contrast to order parameter correlators,
exact analytic results have been known for many decades, both in
equilibrium \cite{n-67,mc} and after a quantum quench \cite{mc}. Here
we summarize these results and point out some interesting features,
which as far as we know have not previously been stressed in the
literature. 

\subsection{The one-point function}
We first consider the one-point function, \emph{i.e.} the
magnetization in the $z$ direction, which is \cite{mc}
\be
\fl\qquad \braket{\sigma_l^z}
=\braket{i a_l^y(t) a_l^x(t)}=-\int_0^\pi\frac{\mathrm d k}{\pi}
\Bigl(\cos\theta_k\cos\Delta_k+\sin\theta_k\sin\Delta_k\cos(2\veps_h(k)
t)\Bigr)\, ,
\label{1pointz}
\ee
where $\veps_h(k)$, $\theta_k$, $\Delta_k$ and defined in \fr{Hbog},
\fr{Bangle} and \fr{cosDeltak} respectively.
Unlike the order parameter, the transverse magnetization has a 
finite value in the stationary state
\be
\lim_{t\to\infty}\braket{\sigma_l^z} =
-\int_0^\pi\frac{\mathrm d k}{\pi}
\cos\theta_k\cos\Delta_k\, . 
\label{tinftyt}
\ee
The result \fr{tinftyt} agrees with the GGE prediction.
The late-time limit of \fr{1pointz} can be determined by a stationary
phase approximation. The leading contributions arise from the saddle
points at $k=0$ and $k=\pi$ and a straightforward calculation gives
\begin{eqnarray}
\fl\quad
\braket{\sigma_l^z}\Big|_{J t\gg 1} &=
-\int_0^\pi\frac{\mathrm d k}{\pi} \cos\theta_k\cos\Delta_k\\
\fl&+\frac{h_0-h}{4\sqrt{\pi}(2 h J t)^{\frac{3}{2}}}\Bigl[\frac{\sin(4 J |1-h|t+\pi/4)}{\sqrt{|1-h|}|1-h_0|}-\frac{\sin(4 J (1+h)t-\pi/4)}{\sqrt{1+h}(1+h_0)}\Bigr]+O((J t)^{-\frac{5}{2}})\, .
\end{eqnarray}

\subsection{The two-point function}
The connected transverse two-point
correlator is expressed in terms of the Majorana fermions as (see
Eq. (\ref{corrferm})) 
\bea
\rho_c^{zz}(\ell,t)&=
\braket{\sigma_l^z\sigma_{l+\ell}^z}-\braket{\sigma^z_l}^2\nn
&=\braket{a_1^y a_{\ell+1}^y}\braket{a_1^x a_{\ell+1}^x}-\braket{a_1^y
  a_{\ell+1}^x}\braket{a_1^x a_{\ell+1}^y}
=f_{-\ell}f_\ell-g_{\ell+1}g_{1-\ell}\,.  
\eea
Using  \Eref{eq:fg}, $\rho_c^{zz}(\ell,t)$ can be written as a double
integral in momentum space 
\bea
\fl\quad
\label{eq:doubletrans}
\rho_c^{zz}(\ell,t)
&=\int_{-\pi}^\pi\frac{\mathrm d k_1\mathrm d k_2}{(2\pi)^2}
e^{i \ell (k_1-k_2)}\Bigg\{\prod_{j=1}^2\sin\Delta_{k_j}
\sin\big(2\veps_h(k_j)t\big)\nn
\fl&\qquad\qquad\qquad\qquad-
\prod_{j=1}^2e^{i\theta_{k_j}}\Big[\cos\Delta_{k_j}-i\sin\Delta_{k_j}\cos\big(2\veps_h(k_j)t\big)\Big]\Bigg\}\, .
\eea
where $\theta_k$, $\Delta_k$ and $\veps_h(k)$ are given by \fr{Bangle},
\fr{cosDeltak} and \fr{Hbog} respectively.

\subsubsection{The infinite time limit and the GGE.}
In the limit $t\to\infty$ at fixed, finite
$\ell$ all oscillating terms in this integral average to zero and we
are left with 
\be\label{eq:rhoz}
\rho_c^{zz}(\ell,\infty)=-\int_{-\pi}^\pi \frac{\mathrm d k}{2\pi}e^{i
  \ell k}e^{i \theta_k}\cos\Delta_k\  
\int_{-\pi}^\pi \frac{\mathrm d p}{2\pi}e^{i \ell p}e^{-i \theta_p}\cos\Delta_p\, .
\ee
This resembles the result for the finite temperature connected 2-point
function obtained in \cite{mc}
\be
\fl\qquad
\langle\!\langle{\sigma^z_l\sigma^z_{l+\ell}}\rangle\!\rangle-
\langle\!\langle{\sigma^z_{l}}\rangle\!\rangle^2
=-\int_{-\pi}^\pi \frac{\mathrm d k}{2\pi}e^{i \ell k}e^{i
  \theta_k}\tanh\Bigl(\frac{\beta\veps_h(k)}{2}\Bigr)\
\int_{-\pi}^\pi \frac{\mathrm d p}{2\pi}e^{i \ell p}e^{-i \theta_p}\tanh\Bigl(\frac{\beta\veps_h(p)}{2}\Bigr)\, .
\label{zzT}
\ee
Here $\langle\!\langle{\cal O}\rangle\!\rangle$ denotes the thermal
equilibrium expectation value at temperature $1/\beta$.
Replacing $\beta$ with a mode dependent inverse temperature as in
(\ref{betak}),
\be
\beta\veps_h(k)\longrightarrow 
\beta_k\veps_h(k)=2\ {\rm arctanh}\big(\cos(\Delta_k)\big)\ ,
\ee
reduces \fr{zzT} to \fr{eq:rhoz}. This again confirms the notion that
the GGE can be thought of in terms of a mode-dependent temperature.
In order to determine the large$-\ell$ behaviour of
$\rho^{zz}_c(\ell,\infty)$ we change the integration variable in
(\ref{eq:rhoz}) to $z=e^{i k}$. This results in a contour integral
along the unit circle $C$
\bea
\label{eq:rhoz1}
\rho^{zz}_c(\ell,\infty)=-\frac{(h+h_0)^2}{4 h h_0}I_+ I_-\, ,\nn
I_\pm=\oint_C\frac{\mathrm d z}{2\pi
  i}\frac{z^{\ell-1{\pm 1}/{2}}}{\sqrt{(z-1/h_0)(h_0-z)}}\frac{(z-1/h_1)(h_1-z)}{\pm(z-h^{\mp
  1})}\, ,
\eea
where $h_1$ is given in (\ref{eq:h1}). Inside the unit circle there is
a branch cut connecting $0$ with $\min\{h_0,1/h_0\}$. In addition
there is a pole at $\min\{h,1/h\}$ in $I_+$ if $h>1$ and in $I_-$ if $h<1$.
The leading contributions to the integral arise from the pole and the
vicinity of the branch point at $\min\{h_0,1/h_0\}$. Taking these into
account (and assuming that $h_0h\neq 1$) we obtain
\be
\fl\qquad
\rho^{zz}_c(\ell,\infty)=
\cases{
{\cal C}^{zz}_1\ \frac{e^{-2|\ln h_0|\ell}}{\ell}\big(1+O(\ell^{-1})\big)
&if $|\ln h|>|\ln h_0|$\, ,\cr
{\cal C}^{zz}_2\ \frac{e^{-(|\ln h|+|\ln
    h_0|)\ell}}{\sqrt{\ell}}\big(1+O(\ell^{-1})\big)&if $|\ln h_0|>|\ln h|$.\cr}
\ee 
Here the amplitudes are
\bea
{\cal C}^{zz}_1&=
\frac{|h_0-1/h_0|}{4\pi}\frac{h-h_0}{1-h h_0}\ ,\nn
{\cal C}^{zz}_2&=
\frac{(h-1/h)\sqrt{|h_0-1/h_0|}(h_0-h)}{8\sqrt{\pi}
  h}\sqrt{\frac{h_0-h}{h_0(h h_0-1)}}\frac{e^{\mathrm{sgn}(\ln h)|\ln
    h_0|/2}}{\sinh\frac{|\ln h|+|\ln h_0|}{2}} \ .
\end{eqnarray}
For quenches such that $hh_0=1$ the result is instead
\be
\fl\qquad\rho^{zz}_c(\ell,\infty)=
-\frac{(h-1/h)^2}{2\pi} e^{-2 |\ln h|\ell}+\ldots\qquad h_0=1/h\, .
\ee
\subsubsection{Time dependence in the space-time scaling regime.}
We now consider the late time behaviour in the space-time scaling
regime, i.e. in an asymptotic expansion around the limit $\ell,t\to
\infty$ at fixed finite ratio $v=\ell/2t$. As $\ell\propto t$ appears
in \fr{eq:doubletrans} linearly in the phases (albeit in different
combinations), the asymptotic power law behaviour can be determined by
a stationary phase approximation. The calculation is relatively
simple, but the resulting formulas are long and not very
illuminating. The general form of the answer is
\be
\rho^{zz}_c(\ell,t)\simeq \frac{1}{t} \sum_i  \alpha_i(v) \cos(\omega_i(v) t+\varphi_i(v))\,,
\label{rhozsum}
\ee
where the sum runs over the various saddle points. 
The leading power-law $t^{-1}$ comes simply from the stationary phase,
but the calculations of the functions $\alpha_i$, $\varphi_i$ and
$\omega_i$ are quite tedious. For this reason we focus on the
non-oscillating contribution with $\omega_0(v)=0$, which we denote by
$\rho_0^{zz}(\ell,t)$. This zero-frequency piece arises
from terms, in which the phase in \Eref{eq:doubletrans} vanishes at
the stationary points, i.e. from 
\be\fl\qquad
\int\limits_{-\pi}^\pi\frac{\mathrm d k_1\mathrm d k_2}{8\pi^2}e^{2i v
  t (k_1-k_2)}
\sin\Delta_{k_1}\sin\Delta_{k_2}\cos\big(2[\veps_h(k_1)-\veps_h(k_2)]t\big)
\bigl(1+e^{i(\theta_{k_1}+\theta_{k_2})}\bigr)\,
. 
\label{spleft}
\ee
The large-$t$ behaviour of this double integral can be extracted by a
two-dimensional stationary phase approximation. 
Its non-oscillating part (characterized by $k_1=k_2$) for $h\neq 1$ is 
\bea\label{eq:averrho}
\rho_0^{zz}(\ell,t)&\simeq& 
\frac{1}{4t}\int_{0}^\pi\frac{\mathrm d k}{\pi}
\sin^2\Delta_k\cos^2\theta_k\ \delta(\veps'_h(k)-v)
\\&=&
\frac{\theta_H(v_{\rm max}-v)}{4\pi t}\Bigl[\frac{\sin^2\Delta_{k_v}\cos^2\theta_{k_v}}{|\eps^{\prime\prime}(k_v)|}+\frac{\sin^2\Delta_{\tilde k_v}\cos^2\theta_{\tilde k_v}}{|\eps^{\prime\prime}(\tilde k_v)|}\Bigr]\, ,
\label{2ndline}
\eea
where $k_v$ and $\tilde{k}_v$ are the solutions of $\veps'_h(k)=v$. In
the case $h=1$ there is only a single term in \fr{2ndline} with $k_v=
2{\rm arccos}(v/2J)$. The following remarks are in order
\begin{enumerate}
\item{}
For $v>v_{\rm max}$ the connected two-point
function is exponentially small in $vt=\ell\gg 1$, in accordance
with expectations based on causality \cite{cc-06}. 
\item{}
In contrast to the order parameter two-point function
$\rho^{xx}_c(\ell,t)$ studied in paper I, for given $t$ and $\ell$
only elementary excitations with velocity exactly equal to $v=\pm
\ell/2t$ contribute to the asymptotic behaviour of $\rho^{zz}_c(\ell,t)$.
This is a consequence of the transverse spins being fermion bilinears.
\item{} In the limit $v\to0$ \fr{eq:averrho} becomes
($\sin\Delta_k\propto v$, while $\cos^2\theta_k$ and $\veps''_h(k)$
  are finite at the saddle points),
\be
\rho_0^{zz}(\ell,t)\propto \ell^2/t^3\ .
\ee
\end{enumerate}

\section{Conclusions}
In this work we have derived analytic expressions for the stationary
behaviour of the reduced density matrix as well as one and two point
functions in the transverse field Ising chain after a sudden quench of
the magnetic field. We have shown that local observables in the
infinite time limit are described by an appropriately defined
\emph{generalized Gibbs ensemble}. We have furthermore analyzed the
approach in time to the stationary state and addressed the question,
how long it takes for the GGE to reveal itself for a given observable.
We found that in general this time grows exponentially with the
spatial extent of the observable considered, e.g. $t\propto
e^{\alpha\ell}$ for a two-point function with spatial separation $\ell$.
Our analysis can be straightforwardly generalized to quantum quenches
in other models with free fermionic spectrum such as the spin-1/2 XY
chain in a magnetic field and we expect qualitatively similar 
behaviour. In contrast, the stationary behaviour in interacting
integrable models such as the $\delta$-function Bose gas is not
amenable to treatment by the methods employed here and one needs to
proceed along different paths \cite{ksci-12}.
\ack
We thank John Cardy, Jean-Sebasti\'en Caux, Robert Konik, Dirk
Schuricht and Alessandro Silva for helpful discussions. This work was
supported by the EPSRC under grant EP/I032487/1 (FHLE and MF) and the
ERC under the Starting Grant n. 279391 EDEQS (PC). 
This work has been partly done when the authors were guests of the Galileo Galilei 
Institute in Florence whose hospitality is kindly acknowledged.

\appendix
\setcounter{section}{0}

\section{Asymptotic behaviour of determinants of Toeplitz matrix: Szeg\"o Lemma, Fisher Hartwig conjecture and generalization}
\label{szapp}

In this appendix we summarize results on Toeplitz determinants, 
that are used in the main part of the paper. 
Let $\big(T_\ell\big)_{l n}= t_{l-n}$ be a $\ell\times \ell$ Toeplitz matrix with
symbol $t(e^{i k})$, \emph{i.e.} 
\be\label{eq:Fourier}
 t_n\equiv \int_0^{2\pi}\frac{\mathrm d k}{2\pi}t(e^{i k})e^{- i n k}\mathrm d k\, .
\ee
The behaviour of the determinant of $T$ for asymptotically large
$\ell$ depends on the analytic properties of the symbol $t(e^{ik})$.

\subsection{Smooth symbol: The strong Szeg\"o Lemma}

If the symbol $t(z)$ is a {\it nonzero} continuous function of $z$ on the integration countour, 
with winding number zero about the origin, 
the function $\log t(z)$ admits the Fourier (Laurent) expansion
\be
\log t(z)=\sum_{q=-\infty}^\infty (\log t)_q z^{q},\qquad (\log t)_q=\int_{0}^{2\pi}\frac{\mathrm d k}{2\pi} \log t(e^{i k})e^{- i k q}\,.
\label{Vz}
\ee
The asymptotic behaviour of the determinant is then given by the
strong Sz\"ego limit theorem \cite{FH-gen}, which states that
\bea
\label{eq:Szego}
\det {[T_\ell]}&=& E_{[t]}\ e^{\ell (\log t)_0}
\big(1+{\cal O}(\ell^{1-2\beta})\big)\ ,
\eea
where
\be
E_{[t]}=  \exp\Bigl[{\sum_{q\geq 1}q (\log t)_q (\log t)_{-q}}
  \Bigr]\, .
\ee
A convenient alternative form of this result is
\be
\ln \det{[T_\ell]}=\ell \int_0^{2\pi}\frac{\mathrm d k}{2\pi}\ln t(e^{i k})+\sum_{q\geq 1}q (\log t)_q (\log t)_{-q}
+O(\ell^{1-2\beta})\, .
\label{app1}
\ee
The exponent $\beta$ characterizing the subleading corrections
is determined by the analytic properties of $t(e^{i k})$. 
The integer part of $\beta$ is equal to the number of continuous
derivatives of $t(e^{i  k})$. If the symbol is infinitely differentiable
the corrections fall off faster than any power of $\ell$.

\subsection{Symbols with zeroes and singularities but zero winding 
number: The Fisher Hartwig conjecture}

One of the many generalizations of the Sz\"ego theorem 
is the Fisher Hartwig conjecture. It applies to cases in which
the symbol $t(e^{ik})$ has zeroes and/or discontinuities and can be
expressed in the form
\be\label{eq:symbolform}
t(e^{i k})=t_0(e^{i k})\prod_{r=0}^{R}e^{i \beta_r(k-k_r-\pi\mathrm{sgn}(k-k_r))}(2-2\cos(k-k_r))^{\alpha_r}\, .
\ee
Here $R$ is an integer, $\alpha_r$, $\beta_r$, and 
$0=k_0<k_1<\cdots<k_R<2\pi$ are constants characterizing the location
and nature of singularities and zeroes, and $t_0(e^{i k})$ is a smooth
nonvanishing function with winding number zero. 
In terms of the variable $z=e^{ik}$ the symbol is 
\be\label{eq:symbolform1}
t(z)=t_0(z) z^{\sum_{j=0}^R \beta_j}\prod_{j=0}^R\Bigl(2-\frac{z_j}{z}-\frac{z}{z_j}\Bigr)^{\alpha_j} g_{\beta_j}(z)z_j^{-\beta_j}\,,
\ee
where $z_j=e^{i k_j}$, $j=0,\dots, R$ and
\be
g_{\beta_j}(z)=\left\{\begin{array}{ll}
e^{i\pi\beta_j}&0\leq \arg z<k_j\\
e^{-i\pi\beta_j}&k_j\leq \arg z<2\pi.
\end{array}
\right.
\ee
We follow conventions in which $z_0=1$ is always included in the set
$\{z_j\}$, while for all other $z_{j}$ we must have either $\alpha_j
\neq 0$, $\beta_j\neq 0$ or both. Following the notations of
Ref. \cite{FH-DIK} we introduce the Wiener-Hopf factorization of
$t_0(z)$ as
\bea
t_0(z)&=b_+(z)e^{(\log t_0)_0}b_-(z),\nn
b_+(z)&=e^{\sum_{q\geq 1}(\log t)_q z^q},\quad b_-(z)=e^{\sum_{q\leq -1}(\log t)_q z^q}\, .
\label{WHf}
\eea
Here the constant piece $e^{(\log t_0)_0}$ has been fixed through the
requirement $(\ln b_+)_0=0$. When $\mathrm{Re}[\alpha_j]>-1/2$ and for
$\alpha_j\pm\beta_j\notin\mathbb{Z}^-$, the Fisher-Hartwig conjecture
(which under the above conditions has been proved \cite{Ehr}) gives
the asymptotic behaviour of the determinant 
\begin{eqnarray}\label{eq:FH}
\fl\quad
\det{\Bigl[T_\ell\Bigr]}\sim& E_{[t_0]}\ \e^{\ell (\log t_0)_0}\
\left(\prod_{j=0}^R
\left[b_+(z_j)\right]^{-\alpha_j+\beta_j}
\left[b_-(z_j)\right]^{-\alpha_j-\beta_j}\right)\ell^{\sum_{j=0}^R
   (\alpha_j^2-\beta_j^2)}\nonumber\\ 
\fl&
\times \left[
\prod_{0\leq j<k\leq R}|z_j-z_k|^{2(\beta_j\beta_k-\alpha_j\alpha_k)}
\Bigl(\frac{z_k}{z_j
  e^{i\pi}}\Bigr)^{\alpha_j\beta_k-\alpha_k\beta_j}\right]\nn
\fl&\qquad\times\left[
\prod_{j=0}^R\frac{G(1+\alpha_j+\beta_j)G(1+\alpha_j-\beta_j)}
{G(1+2\alpha_j)}\right]\, .
\end{eqnarray} 
Here the functional $E_{[t]}$ is given in (\ref{eq:Szego}) and $G(x)$
is the Barnes G function. In the special case
$R=0$, $\alpha_0=\beta_0=0$ the strong Sz\"ego lemma applies and the
above formula reduces to (\ref{eq:Szego}).
In cases where $\alpha_j$ and $\beta_j$ don't satisfy the above
requirements, the large-$\ell$ asymptotic behaviour can be determined
from the so-called generalized Fisher-Hartwig conjecture. Now 
there are in general several different \emph{representations} of the symbol 
in the form (\ref{eq:symbolform}), i.e. there are several possible
choices of the parameters $\{\beta_r\}$ in \fr{eq:symbolform}.
The asymptotic behaviour of the determinant is then given as a sum
over all inequivalent representations
\be
\det{\left[T_\ell\right]}\simeq \sum_{\{\beta_r\}_{\rm ineq}}
\det{\left[T_\ell\right]}_{\{\beta_r\}}\,, 
\ee 
where $ \det{\left[T_\ell\right]}_{\{\beta_r\}}$ denotes the expression
(\ref{eq:FH}) for a given set $\{\beta_r\}$. In the cases encountered
in the main part of our paper we identify the representations
giving rise to the leading asymptotics and quote only their contribution.

\subsection{The case of a symbol with non-zero winding number}

A generalization of the Sz\"ego theorem for symbols with nonzero
winding number does exist \cite{sviaj}. If the symbol has negative
winding number, \emph{i.e.} $\exists \kappa\in\mathbb{N}^+$ such that

\be
a(e^{i k})\equiv (-1)^\kappa e^{i \kappa k}t(e^{i k})
\ee
has winding number zero about the origin, the large-$\ell$ asymptotics
of the determinant is given by \cite{FH-gen,FH-DIK,sviaj}
\begin{eqnarray}\label{eq:poswind}
\fl \qquad \det \left[T_\ell\right]=E_{[a]}\ 
\exp\Bigl(\ell \int_0^{2\pi}\frac{\mathrm d k}{2\pi}\ln [a(e^{i k})]\Bigr)
\Bigl(\det\left[\widetilde{T}_\kappa
\right]+{\cal O}\big(\ell^{-3 \beta}\big)\Bigr)
(1+{\cal O}\big(\ell^{1-2\beta})\big)\, ,
\end{eqnarray}
where $E_{[a]}$ is given by (\ref{eq:Szego}) and
$\widetilde{T}_\kappa$ is a $\kappa\times\kappa$ Toeplitz
matrix with elements
\be
\left(\widetilde{T}_\kappa\right)_{ln}=
\int_{-\pi}^\pi\frac{dk}{2\pi}\ e^{-i(l-n)k}
e^{-i \ell k}\frac{a_-(e^{i k})}{a_+(e^{i k})}\ ,
\ee
where
\be
a(e^{i k})=a_+(e^{i k})e^{(\log a)_0}a_-(e^{i k}) \qquad a_\pm(e^{i k})=\exp\Bigl[\sum_{j=1}^\infty(\ln a)_{\pm j}e^{\pm i j k }\Bigr]\, .
\ee
As before we use notations where $(\ln a)_{j}$ are the Fourier
coefficients of $\ln a$. For our purposes we only need to consider
the case $\kappa=1$, in which \fr{eq:poswind} takes the simpler form
\begin{eqnarray}
\fl\qquad\qquad\det T_\ell\sim E_{[a]}\ e^{\ell (\log
  a)_0}\int_0^{2\pi}\frac{\mathrm d k}{2\pi}e^{-i \ell k}\ \frac{a_-(e^{i k})}{
a_+(e^{i k}) }
+\dots\, .
\label{FHwnm1}
\end{eqnarray}
At large $\ell$ we have $\det[T_\ell]\propto e^{\ell/\xi}$, where
\be
\xi^{-1}=\lim_{\ell\rightarrow\infty}\frac{\det T_\ell}{\ell} =
(\log
a)_0+\lim_{\ell\rightarrow\infty}\frac{1}{\ell}\ln\int_0^{2\pi}\frac{\mathrm
  d k}{2\pi}e^{-i \ell k}\ \frac{a_-(e^{i k})}{a_+(e^{i k})}\,.
\ee

The case of a symbol with positive winding number can be obtained
directly from (\ref{eq:poswind}) and (\ref{FHwnm1}) by noting that the
transpose of a Toeplitz matrix is another Toeplitz matrix with a symbol
of opposite winding number. 

\section*{References}


\end{document}